\documentclass[aps,prd,groupedaddress,preprint,eqsecnum,nofootinbib,showpacs]{revtex4}
\usepackage{graphicx,epsf,amssymb,amsbsy,amsfonts,amssymb,amsmath}
\usepackage[dvips]{color}
\usepackage{ifthen}

\newif\ifdraft
\drafttrue
\newif\ifpreprint
\preprinttrue

\def\sect#1{section~{\ref{#1}}}
\def\fig#1{fig.~{\ref{#1}}}

\def\figs#1#2{figs.~{\ref{#1}} and {\ref{#2}}}

\def\tab#1{table~{\ref{#1}}}

\def\tabs#1#2{tables~{\ref{#1}} and {\ref{#2}}}

\def\spa#1.#2{\left\langle#1\,#2\right\rangle}
\def\spb#1.#2{\left[#1\,#2\right]}
\def\spash#1.#2{\spa{\smash{#1}}.{\smash{#2}}}
\def\spbsh#1.#2{\spb{\smash{#1}}.{\smash{#2}}}
\def\sand#1.#2.#3{%
\left\langle\smash{#1}{\vphantom1}^{-}\right|{#2}%
\left|\smash{#3}{\vphantom1}^{-}\right\rangle}
\def\sandpp#1.#2.#3{%
\left\langle\smash{#1}{\vphantom1}^{+}\right|{#2}%
\left|\smash{#3}{\vphantom1}^{+}\right\rangle}
\def\sandpm#1.#2.#3{%
\left\langle\smash{#1}{\vphantom1}^{+}\right|{#2}%
\left|\smash{#3}{\vphantom1}^{-}\right\rangle}
\def\sandmp#1.#2.#3{%
\left\langle\smash{#1}{\vphantom1}^{-}\right|{#2}%
\left|\smash{#3}{\vphantom1}^{+}\right\rangle}

\def\twoloop{{2 \mbox{-} \rm loop}}

\def\tree{{\rm tree}}

\def\Tr{\, {\rm Tr}}

\def\eps{\epsilon}
\def\e{\epsilon}
\def\ep{\epsilon}

\def\nn{\nonumber}

\def\eqn#1{eq.~(\ref{#1})}

\def\eqns#1#2{eqs.~(\ref{#1}) and~(\ref{#2})}

\def\Neqfour{{{\cal N}=4}}
\def\NeqFour{{{\cal N}=4}}
\def\Neqeight{{{\cal N}=8}}
\def\NeqEight{{{\cal N}=8}}
\def\NeqOne{{{\cal N}=1}}

\def\f{\tilde f}
\def\be{\begin{equation}}
\def\ee{\end{equation}}
\def\bea{\begin{eqnarray}}
\def\eea{\end{eqnarray}}
\def\ba{\begin{eqnarray}}
\def\ea{\end{eqnarray}}

\def\tree{{\rm tree}}

\def\threeloop{{\rm 3\hbox{-}loop}}
\def\Ord{{\cal O}}

\newbox\charbox
\newbox\slabox
\def\s#1{{      
        \setbox\charbox=\hbox{$#1$}
        \setbox\slabox=\hbox{$/$}
        \dimen\charbox=\ht\slabox
        \advance\dimen\charbox by -\dp\slabox
        \advance\dimen\charbox by -\ht\charbox
        \advance\dimen\charbox by \dp\charbox
        \divide\dimen\charbox by 2
        \raise-\dimen\charbox\hbox to \wd\charbox{\hss/\hss}
        \llap{$#1$} }}

\def\subtractfour#1{\ifthenelse{#1=5}{1}{\ifthenelse{#1=6}{2}
{\ifthenelse{#1=7}{3}{\ifthenelse{#1=8}{4}{\ifthenelse{#1=9}{5}
{\ifthenelse{#1=10}{6}{\ifthenelse{#1=11}{7}{\ifthenelse{#1=12}{8}
{\ifthenelse{#1=13}{9}{\ifthenelse{#1=14}{10}{}}}}}}}}}}}

\def\g#1#2#3{\tilde f^{a_{#1},a_{#2},a_{#3}}}

\begin{document}

\ifpreprint
 UCLA/08/TEP/24 \hfill $\null\hskip 1.0 cm \null$  SLAC--PUB--13361 
\fi

\title{Manifest Ultraviolet Behavior for the Three-Loop Four-Point Amplitude
of $\boldsymbol{{\cal N}=8}$ Supergravity}

\author{Z.~Bern${}^a$, J.~J.~M.~Carrasco${}^a$, L.~J.~Dixon${}^{b}$,
 H.~Johansson${}^a$ and  R.~Roiban${}^c$ }

\affiliation{
${}^a$Department of Physics and Astronomy, UCLA, Los Angeles, CA
90095-1547, USA  \\
${}^b$Stanford Linear Accelerator Center,
              Stanford University,
             Stanford, CA 94309, USA \\
${}^c$Department of Physics, Pennsylvania State University,
           University Park, PA 16802, USA
}

\date{August, 2008}

\begin{abstract}
Using the method of maximal cuts, we obtain a form of the three-loop
four-point scattering amplitude of $\NeqEight$ supergravity in which
all ultraviolet cancellations are made manifest.  
The Feynman loop integrals that appear have a graphical representation
with only cubic vertices, and numerator factors that are quadratic
in the loop momenta, rather than quartic as in the previous form.
This quadratic behavior reflects cancellations beyond those required
for finiteness, and matches the quadratic behavior of the three-loop 
four-point scattering amplitude in $\NeqFour$ super-Yang-Mills theory.
By direct integration we confirm that no additional cancellations remain
in the $\NeqEight$ supergravity amplitude, thus demonstrating that 
the critical dimension in which the first ultraviolet divergence
occurs at three loops is $D_c=6$.  We also give the
values of the three-loop divergences in $D=7,9,11$.  In addition, we
present the explicitly color-dressed three-loop four-point amplitude of
$\NeqFour$ super-Yang-Mills theory.
\end{abstract}

\pacs{04.65.+e, 11.15.Bt, 11.30.Pb, 11.55.Bq \hspace{1cm}}

\maketitle


\section{Introduction}
\label{Introduction}

Recently, the widely held belief that it is impossible to construct a
consistent four-dimensional point-like theory of quantum gravity has
been called into question for maximally
supersymmetric~\cite{CremmerJuliaScherk} $\NeqEight$ supergravity.  In
particular, an integral representation of the three-loop four-point
amplitude~\cite{GravityThree} in this theory has been obtained
explicitly, which exhibits cancellations beyond those needed for
ultraviolet finiteness.  Moreover, for a class of terms accessible by
isolating one-loop subdiagrams via generalized
unitarity~\cite{GeneralizedUnitarity}, the one-loop ``no-triangle''
property~\cite{OneloopMHVGravity, NoTriangle, NoTriangleSixPt,
NoTriangleKallosh,BjerrumVanhove} -- recently proven in
refs.~\cite{NoTriangleProof, AHCKGravity} -- shows that at least a
subset of these cancellations persist to all loop
orders~\cite{Finite}.  Interestingly, M theory and string theory have
also been used to argue either for the finiteness of $\NeqEight$
supergravity~\cite{DualityArguments}, or that divergences are delayed
through at least nine loops~\cite{Berkovits,GreenII}, though issues
with decoupling towers of massive states~\cite{GOS} may alter these
conclusions.  
If a perturbatively ultraviolet-finite point-like theory of quantum
gravity could be constructed, the underlying mechanism responsible for
the required cancellations would have a profound impact on our
understanding of gravity. 

Over the years supersymmetry has been studied extensively as a
mechanism for delaying the onset of divergences in gravity theories
(see {\it e.g.} refs.~\cite{Supergravity,HoweStelleNew}).  In particular, the
existence, or conjectured existence, of various off-shell superspace
formalisms restricts the form and dimensions of potential
counterterms, leading to bounds on the first permissible loop order at
which an ultraviolet divergence might appear in $\NeqEight$
supergravity.  The precise bound depends on the detailed set of
assumptions.  For example, if an off-shell superspace with ${\cal N} =
6$ supersymmetries manifest were to exist, potential $D=4$ divergences
would be delayed to at least five loops~\cite{HoweStelleNew}, while
the existence of a superspace with ${\cal N}=7$ supersymmetries
manifest would delay the first potential divergence to at least six
loops~\cite{HoweStelleNew}. Similarly, if one were to assume the
existence of a fully covariant off-shell superspace with ${\cal N}=8$
supersymmetries manifest, then the first potential divergence would be
pushed to the seven loop order~\cite{GrisaruSiegel}.  Full superspace
invariants, which could act as potential counterterms, have been
constructed at eight loops, suggesting that a divergence might appear
at this loop order, if it does not appear earlier~\cite{Kallosh}.
This first potential divergence can even be pushed to the nine loop
order, with an additional speculative assumption that all fields
respect ten-dimensional general coordinate
invariance~\cite{KellyPrivate}.  This bound coincides with the one
argued~\cite{GreenII} from the type~II string theory
non-renormalization theorem of Berkovits~\cite{Berkovits}.  Beyond
this order, no purely supersymmetric mechanism has been suggested for
preventing the onset of divergences.  In fact, on dimensional grounds
one can argue that, for any supergravity theory to be ultraviolet
finite to all loop orders, novel cancellations beyond the known
supersymmetric ones {\it must} exist.

Surprisingly, cancellations beyond those implied by naive
loop-momentum power-counting appear to be generic in gravity theories,
as suggested by the one-loop study of ref.~\cite{NoTri}.  This
reference demonstrated that these novel one-loop cancellations are
directly connected to the remarkably good high-energy behavior of
gravity tree amplitudes under the complex deformations used to prove
on-shell recursion relations in gravity~\cite{BCFRecursion,BCFW,
GravityRecursion, CachazoLargez, AHCKGravity, EFKRecursion}.
Recently~\cite{AHK,CheungGravity}, these tree-level properties have
been understood in terms of a space-like gauge similar to light-cone
gauge~\cite{SpaceCone}, where an enhanced Lorentz symmetry was shown
to exist under large complex deformations.

For non-supersymmetric theories, these cancellations are
insufficient to render the theory ultraviolet finite.  Indeed,
it is a classical result that gravity coupled to matter generically
diverges at one loop~\cite{tHooftVeltmanGravity, DeserMatter,
DunbarNorridge}.  Pure Einstein gravity does not possess a viable
counterterm at one loop, delaying the divergence to two
loops~\cite{tHooftVeltmanGravity,Kallosh74,vanNWu}.  The
presence of a two-loop divergence in pure Einstein gravity was
established by Goroff and Sagnotti and by van~de~Ven, through direct
computation~\cite{GoroffSagnotti,vandeVen}.

Supersymmetric cancellations can often act on top of any generic
cancellations.  In particular, in the case of $\NeqEight$ supergravity
at one loop, such cancellations combine to cause the vanishing of the
coefficients of all scalar bubble and triangle
integrals~\cite{OneloopMHVGravity, NoTriangle, NoTriangleSixPt,
NoTriangleKallosh,BjerrumVanhove}, proofs of which have been given
recently~\cite{NoTriangleProof,AHCKGravity}.  The one-loop amplitudes
can be expressed solely in terms of box integrals, multiplied by
rational coefficients, exactly as is the case for $\Neqfour$
super-Yang-Mills theory~\cite{UnitarityMethod}.  This one-loop
no-triangle property can be used, via unitarity, to understand a class
of higher-loop cancellations~\cite{Finite}.  However, it does not
account for all of them, because regions where two or more overlapping
loop momenta become large are not covered directly by the one-loop
analysis.  More generally, the picture that emerges at higher loops is
that the excellent ultraviolet behavior found in explicit
calculations~\cite{BDDPR,GravityThree} is due to a combination of
generic cancellations with supersymmetric ones~\cite{Finite,NoTri}.
Interestingly, the absence of bubble integrals in theories with ${\cal
N} \ge 5$ supersymmetry at one loop might lead one to speculate that
such theories may also be finite, if ${\cal N} = 8$ supergravity is
finite~\cite{NoTri}.

To establish the critical dimension where divergences first occur at a
given loop order, we evaluate amplitudes directly.  A general strategy
for obtaining loop amplitudes in gravity theories was first given in
ref.~\cite{BDDPR}, following earlier work in $\NeqFour$
super-Yang-Mills theory at one loop~\cite{UnitarityMethod} and higher
loops~\cite{BRY}.  Using generalized unitarity, at any loop order we
can evaluate scattering amplitudes from products of on-shell tree
amplitudes~\cite{GeneralizedUnitarity}.  In gravity theories we can
then exploit the Kawai-Lewellen-Tye (KLT) relations~\cite{KLT}, which
express gravity tree amplitudes directly in terms of gauge theory tree
amplitudes.  (Subsequent generalizations and other approaches to these
relations may be found in refs.~\cite{KLT2,GravityReview,BEZ,TreeJacobi}.)  The
net effect is that we can map complicated gravity calculations into
substantially simpler gauge theory calculations.  This simplification
is especially important when evaluating generalized unitarity cuts in
$D$ dimensions.

In ref.~\cite{GravityThree} this strategy was used to obtain the
complete three-loop four-point amplitude of $\NeqEight$ supergravity.
The result of this calculation explicitly demonstrated hidden
cancellations beyond those identified in the earlier partial
calculation of ref.~\cite{BDDPR}.  However, the cancellations
took place between different terms of the integral representation
of ref.~\cite{GravityThree}.   Each of the nine contributing ``parent''
integrals had a graphical representation with only cubic vertices, 
and numerator factors that were quartic in the loop momenta.  
(In general, a parent integral refers to an integral with the 
maximal number of propagators allowed in a given amplitude.
The graphs for such integrals have only cubic vertices.)
One of the main purposes of the present paper is to provide an 
improved form of the amplitude in which the true ultraviolet 
behavior is manifest. The new form can be written in terms of the same 
nine parent integrals, but terms have been shuffled between the
different integrals, so that each numerator factor is now quadratic
in the loop momenta.  In this way, no term in the amplitude has a 
worse ultraviolet behavior than does the sum over all terms.

The new representation of the three-loop four-point amplitude is
constructed using the technique of maximal cuts, developed in
ref.~\cite{FiveLoop}.  This technique uses generalized
unitarity~\cite{GeneralizedUnitarity}, starting with the maximum
number of cut propagators and systematically reducing the number of
cut propagators, in order to construct the complete amplitude.  As
observed by Britto, Cachazo and Feng~\cite{BCFGeneralized}, on-shell
massless three-point amplitudes can be defined by analytically
continuing momenta to complex
values~\cite{GoroffSagnotti,WittenTopologicalString}.  Maximal cuts
involve products of three-point tree amplitudes alone, and are the
simplest cuts to evaluate.  Near-maximal cuts, in which one or two of
the maximal-cut propagators have been allowed to go off-shell, are the
next simplest to consider, and so on. We use the near maximal cuts to
fix contact terms that may have been missed by the purely maximal
ones. The maximal-cut technique, by removing cut conditions one by one
from the various maximal cuts, allows us to focus at each stage on a
small subset of contributions to an ansatz for the amplitude.  In this
way, we can efficiently find compact representations of amplitudes
with the desired properties.  The related ``leading-singularity''
technique~\cite{FreddyMaximal} is also based on cutting the maximal
number of propagators.  In this technique, additional hidden
singularities are used.  This technique has been used to compute two-
and three-loop planar $\NeqFour$ super-Yang-Mills
amplitudes~\cite{LeadingSingularityCalcs} with more than four external
states.  An interesting recent conjecture is that the leading
singularities may be sufficient to determine multi-loop amplitudes in
maximally supersymmetric gauge theory and gravity, and that this
property may be linked to the improved ultraviolet behavior of the
theories~\cite{AHCKGravity}.

On-shell gravity and gauge theory amplitudes in four dimensions
possess infrared divergences.  In order to regulate these divergences,
and any potential ultraviolet ones, we work in dimensional
regularization (more specifically, the four-dimensional helicity
scheme~\cite{FDH} related to dimensional reduction~\cite{DimRed}) with
$D=4-2\eps$.  We keep the external states in four dimensions; however,
in the unitarity cuts, the cut loop momenta should be $D$ dimensional.
On the other hand, the maximal-cut method is simplest to implement
initially using four-dimensional instead of $D$-dimensional momenta.
In principle, terms depending solely on the $(-2\eps)$ dimensional
components of momenta may be lost.\footnote{In practice, such terms
appear in theories with less than the maximal amount of supersymmetry,
and in maximally supersymmetric amplitudes with more than four
external states.  However, they have not been found to occur for
four-point amplitudes in $\NeqFour$ super-Yang-Mills theory or
$\NeqEight$ supergravity, although the reason for their absence is still
unclear.}  To confirm that no such terms are missed, we verify that
the results obtained from the maximal-cut method are valid in $D$
dimensions, using the same set of cuts used in
ref.~\cite{GravityThree}.

Recently Nair's on-shell super-space formalism~\cite{Nair} for
$\NeqFour$ super-Yang-Mills MHV amplitudes in four dimensions has been
extended to apply to all amplitudes~\cite{Witten_twistor,
GGK,BEZ,RecentOnShellSuperSpace,CheungGravity,AHCKGravity,EFKRecursion}.
Here we do not use these recent developments to evaluate the
supermultiplet sums appearing in the cuts.  Instead, when evaluating
cuts in four dimensions for $\NeqFour$ super-Yang-Mills amplitudes, we
make use of the observation~\cite{FiveLoop} that four-point amplitudes
are fully determined by considering kinematic choices that force all
or nearly all intermediate states to contain gluons.  In this way we
avoid having to perform any non-trivial sums over super-partners in
the initial construction of an ansatz for the amplitude.  In the
course of verifying the ansatz, we need to evaluate the cuts in $D$
dimensions.  Here we rely on the equivalence of $\NeqFour$
super-Yang-Mills theory to $\NeqOne$ super-Yang-Mills in ten
dimensions, dimensionally reduced to four dimensions.  In ten
dimensions the theory consists of only a gluon and a gluino, greatly
simplifying the supersymmetry bookkeeping.  For the case of
$\NeqEight$ supergravity, we do not need to carry out any explicit
sums over the supermultiplet: the KLT relations allow us to
automatically incorporate all supermultiplet sums directly from the
corresponding super-Yang-Mills results.

Ref.~\cite{GravityThree} demonstrated that the three-loop four-point
amplitude of $\NeqEight$ supergravity is ultraviolet finite for $D<6$.
However, this left unanswered the question of whether there are any
further hidden cancellations which could increase the critical
dimension even further.  Here we address this question by computing
the divergence in $D=6$, using our new representation for the amplitude.
We find that the coefficient of the (logarithmic) divergence is non-zero, 
implying that no further cancellations exist.
This result establishes a nonvanishing coefficient for 
a ``$D^6R^4$'' counterterm for the $D = 6$ version of maximal 
supergravity, whose purely gravitational piece has six derivatives 
acting on a particular combination of four Riemann tensors.
We shall also give values for the divergences in dimensional
regularization in $D=7,9,11$, which are of some interest in studies 
of M theory dualities~\cite{GreenVanhove}.

In this article, we also present the complete three-loop four-point
amplitude of $\NeqFour$ super-Yang-Mills theory, in terms of
a set of integrals dressed by color factors involving nonabelian
structure constants.  (The integrals themselves have already appeared
in ref.~\cite{GravityThree}.)   This amplitude has strong 
infrared divergences in $D=4-2\e$ as $\e\to0$, which begin
at order $1/\e^6$.  The infrared properties of
$\NeqFour$ super-Yang-Mills amplitudes are of some interest, both
in their own right and because of their structural
similarity to those of other gauge theories such as QCD.
Although much of the infrared behavior of gauge theory amplitudes
is well understood through the factorization and exponentiation
of soft and collinear divergences~\cite{IRBehavior}, 
the color-nontrivial soft anomalous dimension matrix~\cite{SoftMatrix}
has only been computed explicitly through two
loops~\cite{AybatDixonSterman}.   At this order the matrix was
found to be proportional to the one-loop matrix, with a proportionality
constant given by the cusp anomalous dimension~\cite{KorchemskyRadyushkin}.
It is natural to conjecture that the same property should hold
at higher orders.

The three-loop soft anomalous dimension matrix is the only unknown 
quantity entering the infrared divergences of the three-loop
four-point amplitude in $\NeqFour$ super-Yang-Mills theory.
It does not appear in the leading-color amplitude~\cite{BDS}.
Thus the $\e$-expansion of the subleading-color terms 
in the complete amplitude would provide a crisp test of the 
proportionality of the matrix at three loops.  
Because soft properties are fairly insensitive to the matter content
of a theory (the states with spin $<1$), if the proportionality holds for 
$\NeqFour$ super-Yang-Mills theory, it is very likely to hold for
a general gauge theory.  The $\e$-expansion of the subleading-color terms 
in the amplitude requires a knowledge of the seven non-planar integrals 
that appear. The two planar integrals are known analytically through
the finite, ${\cal O}(\e^0)$, terms~\cite{SmirnovTripleBox,BDS}.
However, the non-planar integrals pose a more difficult challenge 
and are as yet uncalculated. 

Another application for the $\e$-expansion of the non-planar integrals
would be in the search for potential iterative structures in
subleading-color terms, analogous to those previously found in the planar
amplitudes~\cite{ABDK,BDS,Iterate,LeadingSingularityCalcs}.  

The infrared divergences of one-loop graviton amplitudes were studied
in a classic paper by Weinberg~\cite{Weinberg}. The one-loop divergences
can be exponentiated to give the leading poles in $\e$ at $L$ loops,
$\sim 1/\e^L$.  The infrared behavior is less singular than in gauge
theory because collinear divergences are suppressed, and simpler
because no color matrices appear. However, there are still open
questions about how subleading poles in $\e$ behave.  At two loops,
there has been progress recently in showing how these poles can be
iterated for the four-point $\NeqEight$ supergravity
amplitude~\cite{GravityIR}.  Explicit expressions for the nonplanar
three-loop integrals would help check how this iteration of singular
terms continues to higher order.  It could also be used to search for
any potential pattern of iteration for the finite
terms~\cite{GravityIR}.

It is well known that $\NeqEight$ supergravity contains a non-compact
$E_{7(7)}$ duality symmetry~\cite{CremmerJuliaScherk,E7Original}.
Its explicit action on fields in both light-cone gauge and covariant versions
of the Lagrangian has been studied recently~\cite{E7Recent}.
The non-trivial constraints it imposes on the emission of soft scalars 
at tree level have also been explored~\cite{AHCKGravity}. 
However, there is probably still more to learn about how $E_{7(7)}$ 
constrains multi-loop amplitudes.  Some insight might be provided
by the new form of the three-loop amplitude that we give here.

This paper is organized as follows.  In \sect{ReviewSection} we review
the results previously obtained at two and three loops in both
$\NeqFour$ super-Yang-Mills and $\NeqEight$
supergravity~\cite{GravityThree}. In \sect{ColorDressingSection} we
give the full color-dressed $\NeqFour$ amplitude.  Then in
\sect{SuperGravityResultsSection} we present a form of the $\NeqEight$
supergravity amplitude in which each term exhibits the same ultraviolet
behavior as the full amplitude.  In \sect{DivergencesSection} we show that the
three-loop four-point amplitude diverges logarithmically in $D=6$,
so there there are no further hidden cancellations.
In this section we also present the
three-loop divergences in $D=7,9,11$.  We give our conclusions in
\sect{ConclusionSection}.

\section{General strategy and previous three-loop results}
\label{ReviewSection}

\begin{figure}[t]
\centerline{\epsfxsize 6.0 truein \epsfbox{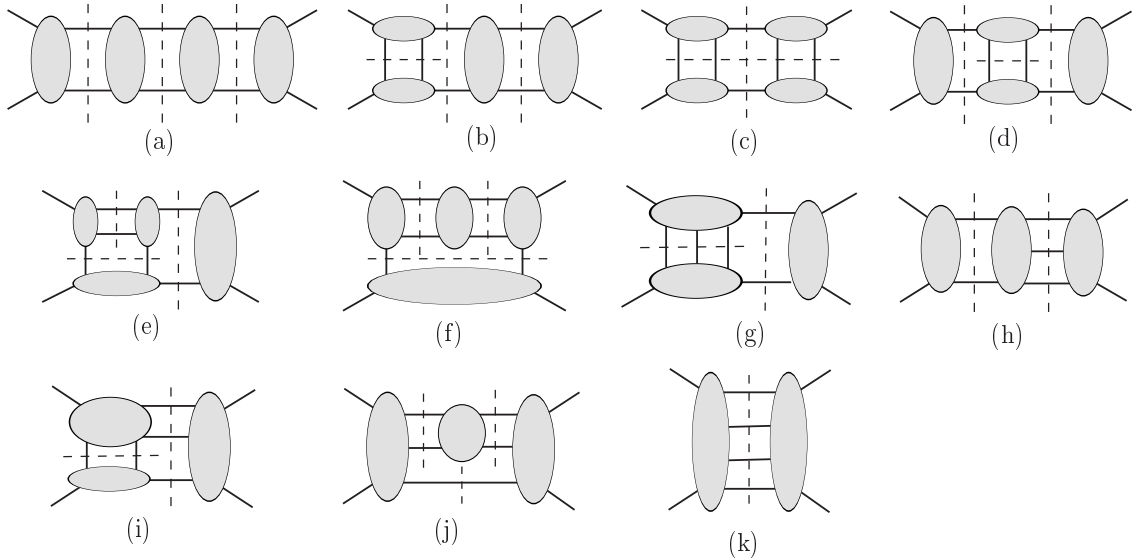}}
\caption[a]{\small A sufficient set of generalized cuts for
determining any massless three-loop four-point amplitude. Each blob
represents a tree amplitude.  Cuts (a)--(f) are iterated two-particle
cuts.}
\label{RealCutsFigure}
\end{figure}

In ref.~\cite{BDDPR} the two-loop four-point $\NeqEight$ supergravity
amplitude was evaluated.  It was also shown that certain classes of
higher-loop contributions could be evaluated to all loop orders.
These so-called ``iterated two-particle cut contributions'' can
be reduced to trees through successive two-particle cuts. 
The same sewing algebra appears in each two-particle cut,
in essentially the same way as in $\NeqFour$ super-Yang-Mills
theory~\cite{BRY}.   In the case of this theory, examining the
powers of loop momenta in the numerator of the generic iterated 
two-particle cut contribution suggested the finiteness bound~\cite{BDDPR},
\begin{equation}
D < \frac{6}{L} + 4  \hskip 1 cm (L > 1) \,,
\label{SuperYangMillsPowerCount}
\end{equation}
where $D$ is the dimension of space-time and $L$ the loop order.  (The
case of one loop, $L=1$, is special; the amplitudes are ultraviolet
finite for $D<8$, not $D<10$.)  The
bound~(\ref{SuperYangMillsPowerCount}) differs somewhat from earlier
superspace power counting~\cite{HoweStelleYangMills}, although all
bounds confirm the ultraviolet (UV) finiteness of $\NeqFour$ super-Yang-Mills
theory in $D=4$. This bound (\ref{SuperYangMillsPowerCount}) has since
been confirmed to all loop orders~\cite{HoweStelleNew} using ${\cal
N}= \nobreak 3$ harmonic
superspace~\cite{HarmonicSuperspace}. Explicit computations
demonstrate that this bound is saturated through at least four
loops~\cite{BRY,BDDPR,ABDK,Finite}.

In ref.~\cite{BDDPR} the iterated two-particle cuts of 
$\NeqEight$ supergravity amplitudes were analyzed, leading
to the proposal that the four-point $\NeqEight$ supergravity
amplitude should be UV finite for
\begin{equation}
D< \frac{10}{L} + 2  \hskip 1 cm  (L>1)\,.
\label{OldPowerCount}
\end{equation}
(Just as for $\NeqFour$ super-Yang-Mills theory, the one-loop case is special;
$\NeqEight$ supergravity amplitudes are UV finite for $D<8$,
not $D<12$~\cite{NoTriangleProof}.)
The formula~(\ref{OldPowerCount}) implies that in $D=4$ the first potential
divergence can appear at five loops.  This result was supported by studying
cuts with an arbitrary number of intermediate states, 
but restricted to maximally-helicity-violating (MHV) amplitudes
on either side of the cut.  
The formula is also consistent with bounds obtained
by Howe and Stelle~\cite{HoweStelleNew}, assuming the existence of 
an ${\cal N} = 6$ harmonic superspace~\cite{HarmonicSuperspace}.  
However, as mentioned in the introduction, explicit three-loop
computations~\cite{GravityThree} have found cancellations beyond this
bound.

At three loops, the generalized cuts shown in \fig{RealCutsFigure} are
sufficient for constructing four-point amplitudes in any massless
theory, starting from tree amplitudes.  Of the cuts in
\fig{RealCutsFigure}, the iterated two-particle cuts (a) to (f) were
originally evaluated in refs.~\cite{BRY,BDDPR} in $\NeqFour$
super-Yang-Mills and $\NeqEight$ supergravity theories.  The remaining
ones, (g) through (k), were evaluated in ref.~\cite{GravityThree} for
both theories, determining the complete amplitudes.  In general, to
determine a three-loop four-point amplitude we must evaluate the cuts
with distinct labels of external legs.  For gravity or color-dressed
gauge theory amplitudes, the various permutations of legs of the tree
amplitudes composing the cuts are automatically included.  However,
color-ordered gauge theory tree-level partial amplitudes must be
explicitly sewn in non-planar fashion to construct higher-loop
non-planar subleading-color amplitudes.

The generalized-cut method for finding the amplitude is algorithmic: one first
constructs an initial ansatz that reproduces one cut. 
Then, subsequent cuts of the amplitude 
are compared against the corresponding cuts of 
the current ansatz. If any discrepancy is found
in a later cut, it is eliminated by adding to the ansatz
terms that vanish when all the earlier cut conditions are imposed.
At the end of this procedure, an integral representation of the loop
amplitude is obtained with the correct cuts in {\it all} channels.
This result is the complete on-shell amplitude. 

In carrying out this construction, tadpole integrals and
bubble contributions on external legs are dropped.  
Such contributions have no cuts on shell, and their integrals
vanish in dimensionally-regularized amplitudes for massless particles.
One might be concerned about potential UV contributions
from bubble integrals on external legs, as $k_i^2 \rightarrow 0$.
Such terms can appear in gauge theory, and cancel collinear infrared 
singularities according to $1/\e_{\rm UV} + 1/\e_{\rm IR} = 0$.
However, in $\NeqEight$ supergravity amplitudes they can be neglected, 
because they are suppressed by additional powers of $k_i^2 \rightarrow 0$
compared to gauge theory, and also by supersymmetric cancellations.

At one loop, through the level of finite terms,
massless supersymmetric amplitudes in the $\NeqFour$
theory are determined completely by their four-dimensional
cuts~\cite{UnitarityMethod}.  Unfortunately, no such theorem has been
demonstrated at higher loops.  Because we wish to identify the smallest
dimension $D$ for which an ultraviolet divergence occurs, we must ensure
that all results are valid in $D$ dimensions.  Evaluating the cuts in
$D$ dimensions~\cite{DDimUnitarity} makes the calculation
significantly more difficult, because powerful four-dimensional spinor
methods~\cite{SpinorHelicity} can no longer be used.  Some of this
additional complexity is avoided by performing internal-state sums in
terms of the (simpler) on-shell gauge supermultiplet of $D=10,\, {\cal
N}=1$ super-Yang-Mills theory instead of the $D=4,\, \NeqFour$
multiplet. By using the KLT relations~\cite{KLT}, 
which are valid in any number of
dimensions for the field content of maximal supergravity, the
$D$-dimensional cuts of $\NeqFour$ super-Yang-Mills theory can then be
reassembled into those of $\NeqEight$ supergravity~\cite{BDDPR}.

\begin{figure}[t]
\centerline{\epsfxsize 5.5 truein \epsfbox{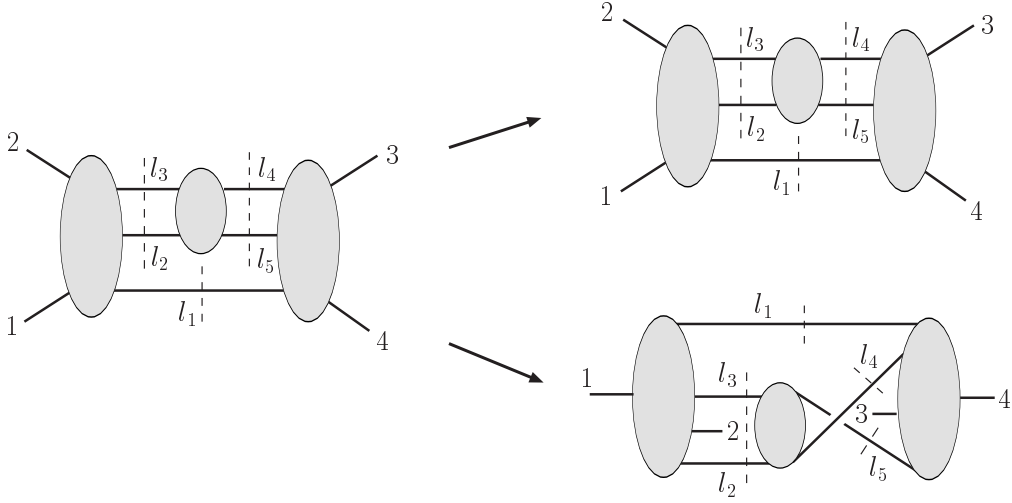}}
\caption[a]{\small The KLT relations allow us to map the gravity 
cuts into sums over pairs of gauge theory cuts.  Here we 
display a pair of gauge theory cuts needed to evaluate the
gravity cut shown in \fig{RealCutsFigure}(j).  The remaining
pairs are obtained by permuting the external legs $1 \leftrightarrow 2$
and $3 \leftrightarrow 4$.}
\label{KLTCutsFigure}
\end{figure}

In general, the KLT relations give us a rather efficient means of
evaluating gravity cuts, since gauge theory amplitudes are generally
simpler to evaluate than gravity amplitudes. As an example, 
consider the cut in \fig{RealCutsFigure}(j) for $\NeqEight$ supergravity,
\begin{equation}
C^{\rm (j)}_{\NeqEight}  = \sum_{\NeqEight\ \rm states} 
M_5^\tree(1, 2, l_3,l_2,l_1) \times  
M_4^\tree(-l_2, -l_3, l_4, l_5) \times
M_5^\tree(3, 4, -l_1, -l_5, -l_4)  \,,
\label{RealCutj}
\end{equation}
where the sum runs over all physical states that cross any cut line.
The KLT relations~\cite{KLT,KLT2,GravityReview} for these tree
amplitudes are,
\begin{eqnarray}
&& \hskip - .2 cm 
M_4^\tree(-l_2, -l_3,, l_4, l_5) = 
 -i \, (l_4 + l_5)^2 \,  A_4^\tree(-l_2, -l_3, l_4, l_5) \, 
   A_4^\tree(-l_2, -l_3, l_5, l_4) \,, \nn \\
&& \hskip - .2 cm 
M_5^\tree(1, 2, l_3,l_2,l_1) = 
 i \, (l_1 + k_1)^2 (l_3 + k_2)^2\,
    A_5^\tree(1,2,l_3,l_2,l_1) \, 
    A_5^\tree(1, l_1, l_3, 2, l_2)  + \{ 1 \leftrightarrow 2 \} \,, \nn \\
&&\hskip - .2 cm 
 M_5^\tree(3,4, -l_1, -l_5, -l_4) = 
 i \,  (l_4 - k_3)^2 (l_1 - k_4)^2 \,
\nn \\
&& \null \hskip 4.6 cm 
 \times  A_5^\tree(3, 4, -l_1, -l_5, -l_4) 
  A_5^\tree(3, -l_4, -l_1, 4, -l_5) \, 
 + \{ 3 \leftrightarrow 4 \} \,,
\nn \\
&& \null \hskip 4.6 cm ~~
\label{KLT45}
\end{eqnarray}
where we follow the notation of ref.~\cite{GravityReview}. The
$A_n^\tree$ are color-ordered gauge-theory tree amplitudes, while 
the $M_n^\tree$ are supergravity tree amplitudes, 
with an overall factor of the coupling $(\kappa/2)^{n-2}$ removed.
Inserting the KLT relations into the cut~(\ref{RealCutj}),
we obtain,
\begin{eqnarray}
C^{\rm (j)}_{\NeqEight} &=& 
i \, 
(l_4 + l_5)^2  (l_1 + k_1)^2 (l_3 + k_2)^2  (l_4 - k_3)^2 (l_1 - k_4)^2 \nn\\
&& \null \times 
 \sum_{\NeqFour\ \rm states}  
   A_5^\tree(1, 2, l_3,l_2, l_1)\, A_4^\tree(-l_2, -l_3, l_4, l_5) \,
       A_5^\tree(3, 4, -l_1, -l_5, -l_4) \nn \\
&& \null \times 
\sum_{\NeqFour\ \rm states}  
       A_5^\tree(1, l_1, l_3, 2, l_2)\, A_4^\tree(-l_2, -l_3, l_5, l_4) \,
        A_5^\tree(3,-l_4, -l_1, 4, -l_5)  \nn \\
&& \null \hskip 3 cm 
 + \{ 1 \leftrightarrow 2 \}  + \{ 3 \leftrightarrow 4\}
 + \{ 1 \leftrightarrow 2,\ 3 \leftrightarrow 4\}\,.
\end{eqnarray}
This equation gives the $\NeqEight$ supergravity cut 
directly in terms of products of two $\NeqFour$ super-Yang-Mills cuts.
The relation is depicted in~\fig{KLTCutsFigure}, for
one of the four terms in the sum over external-leg
permutations.  One of the gauge-theory cuts is planar,
while the second is nonplanar.~\footnote{It is possible to use the 
total $S_3$ permutation symmetry of $s_{12}s_{14}A_4^\tree(1,2,3,4)$ to 
partially ``untwist'' the four-point amplitude in the second Yang-Mills cut, 
so as to make manifest its reflection symmetry under
$\{1 \leftrightarrow 4, \, 2 \leftrightarrow 3\}$.}

An important feature of this construction is that, once the sums over
all super-partners are performed in the $\NeqFour$ super-Yang-Mills
cuts, the corresponding super-partner sum in $\NeqEight$ supergravity
follows simply from the KLT relations.  In fact, any simplifications
performed on the gauge theory cuts can be immediately carried over
to gravity cuts.

\begin{figure}[t]
\centerline{\epsfxsize 5.5 truein \epsfbox{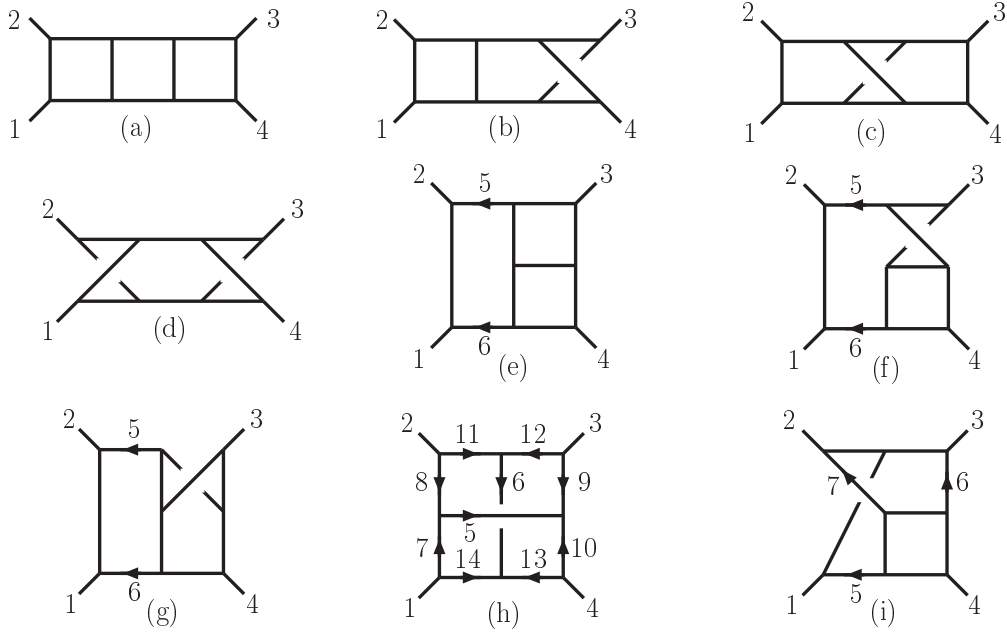}}
\caption[a]{\small The different parent integrals in terms of which 
four-point three-loop amplitudes may be expressed.}
\label{IntegralsThreeLoopFigure}
\end{figure} 

\begin{table*}
\caption{
The numerator factors $N^{(x)}$ for the integrals $I^{(x)}$ in
\fig{IntegralsThreeLoopFigure} for $\NeqFour$ super-Yang-Mills theory. 
The first column labels the integral, the second
column the relative numerator factor. An overall factor of 
$s_{12} s_{14} A_4^\tree$ has been removed.  The invariants
$s_{ij}$ and $\tau_{ij}$ are defined in \eqn{InvariantsDef}.
\label{NumeratorYMTable} }
\vskip .4 cm
\begin{tabular}{||c|c|}
\hline
Integral $I^{(x)}$ & $N^{(x)}$ for $\NeqFour$ Super-Yang-Mills \\
\hline
\hline
(a)--(d) &  $
s_{12}^2
$ 
  \\
\hline
(e)--(g) &  $
s_{12} \,s_{46}
$  \\
\hline
(h)&  $\; 
s_{12} (\tau_{26} + \tau_{36}) +
       s_{14} (\tau_{15} + \tau_{25}) + 
       s_{12} s_{14} 
\;$
\\
\hline
(i) & $\;
s_{12} s_{45} - s_{14} s_{46} -  {1\over 3} (s_{12} - s_{14}) l_7^2
 \;$  \\
\hline
\end{tabular}
\end{table*}

\begin{table*}
\caption{The numerator factors $N^{(x)}$ for the integrals $I^{(x)}$ in
\fig{IntegralsThreeLoopFigure} for $\NeqEight$ supergravity, as determined in
ref.~\cite{GravityThree}.  The first column labels the integral, the
second column the relative numerator factor.  In this form of the amplitude,
individual terms behave worse in the ultraviolet than does the 
sum over all contributions.
\label{NumeratorGravityOldTable} }
\vskip .4 cm
\hskip -.5 cm 
\begin{tabular}{||c|c||}
\hline
Integral $I^{(x)}$ & $N^{(x)}$ for $\NeqEight$ Supergravity  \\
\hline
\hline
(a)--(d) 
& $
\vphantom{\Bigr|}
 [s_{12}^2]^2
$ \\
\hline
(e)--(g) & $\vphantom{\Bigr|}
 [s_{12} \, s_{46}]^2
$  \\
\hline
(h) & $\;
(s_{12} s_{89} + s_{14} s_{11,14} - s_{12} s_{14})^2
          - s_{12}^2 (2 (s_{89} - s_{14}) + l_6^2 ) l_6^2 
          - s_{14}^2 (2 (s_{11,14} - s_{12}) + l_5^2)  l_5^2 
         \; $  \\
    &  $ \; \null
      - s_{12}^2 (2 l_8^2 l_{10}^2 + 2 l_7^2 l_9^2 + l_8^2 l_7^2 
        + l_9^2 l_{10}^2)
       \vphantom{\bigl(A^A_A\bigr) }
 - s_{14}^2 (2 l_{11}^2 l_{13}^2 + 2 l_{12}^2 l_{14}^2 +
            l_{11}^2 l_{12}^2 + l_{13}^2 l_{14}^2 )
           + 2 s_{12} s_{14} l_5^2 l_6^2\;
         $
           \\
\hline
(i) &  $
(s_{12} s_{45} - s_{14} s_{46} )^2 
  \vphantom{\bigl|_{A_A}} \null - ( s_{12}^2  s_{45} + s_{14}^2 s_{46} 
      + {1\over3} s_{12} s_{13} s_{14}) l_7^2
 $ \\
\hline
\end{tabular}
\end{table*}

In the calculation of ref.~\cite{GravityThree}, cut (i) 
in \fig{RealCutsFigure} was used as the starting point.  
In fact, this cut detects all terms in the
complete three-loop four-point amplitude for $\NeqFour$
super-Yang-Mills theory, and almost all terms for the case of
$\NeqEight$ supergravity.  The result of matching cut (i) with an
ansatz for the $\NeqFour$ super-Yang-Mills amplitude, and then
checking the result on the other cuts in \fig{RealCutsFigure}, is that
the amplitude can be expressed as a linear combination of the
parent integrals shown in \fig{IntegralsThreeLoopFigure}, with
numerator factors $N^{(x)}$ given in \tab{NumeratorYMTable}.
Similarly, the three-loop four-point amplitude of $\NeqEight$
supergravity can be expressed in terms of the same parent integrals,
but with different numerators. 
The numerators for this case, as determined in
ref.~\cite{GravityThree}, are shown in \tab{NumeratorGravityOldTable}.
In \fig{IntegralsThreeLoopFigure} the (outgoing) momenta of the
external legs are denoted by $k_i$ with $i=1,2,3,4$, while the momenta
of the internal legs are denoted by $l_i$ with $i\ge5$. For
convenience we define
\begin{eqnarray}
&& s_{ij} = (k_i +k_j)^2 \,,    \hskip 1.5 cm 
\tau_{ij} = 2 k_i\cdot k_j\,,   \hskip 1 cm (i,j \le 4) \nn \\
&& s_{ij} = (k_i +l_j)^2 \,,    \hskip 1.5 cm 
\tau_{ij} = 2 k_i\cdot l_j\,,   \hskip 1.1 cm (i \le 4, j\ge5) \nn \\
&& s_{ij} = (l_i +l_j)^2 \,,    \hskip 1.5 cm 
\tau_{ij} = 2 l_i\cdot l_j\,.   \hskip 1.2 cm (i, j\ge5) 
\label{InvariantsDef}
\end{eqnarray}
We have altered the labeling compared to ref.~\cite{GravityThree}, as
a notational convenience, which will allow us to write somewhat more 
compactly the new representation of the $\NeqEight$
supergravity amplitude constructed here.

By definition, the integrals composing the amplitudes are of the form 
\begin{equation}
I^{(x)}=(-i )^3 
\int  \Biggl[ \prod_{i=1}^3 \frac{d^D q_i}{(2\pi)^D} \Biggr] \, 
 {N^{(x)} \over \prod_{j=5}^{14} l_j^2} \,,
\label{IntegralNormalization}
\end{equation}
where the $q_i$'s are three independent loop momenta, 
the $l_i$'s are the momenta of the propagators of the diagrams,
and the $N^{(x)}$ are the numerator factors appearing in
\tabs{NumeratorYMTable}{NumeratorGravityOldTable}. 
For example, the contribution of diagram (e) of
\fig{IntegralsThreeLoopFigure} to the $\NeqEight$ supergravity
amplitude is found by combining its propagators with the numerator
$N^{\rm (e)}= [s_{12} s_{46}]^2 \equiv [s_{12} (k_4+l_6)^2]^2$ given in
\tab{NumeratorGravityOldTable}; the result is the integral
\begin{eqnarray}
I^{\rm (e)} & = &
(-i )^3\int \frac{d^Dl_6 }{(2\pi)^D} 
\int \frac{d^Dl_7}{(2\pi)^D}   \int \frac{d^Dl_8}{(2\pi)^D} 
{ [s_{12} (k_4+l_6)^2]^2 \over 
 l_6^2 l_7^2 l_8^2  (l_6 - k_1)^2 (l_6 - k_1 - k_2)^2
 (l_7 - k_4)^2 (l_8 - k_3)^2 } \nn \\
&& \hskip 4 cm 
\times{1\over 
  (l_6+l_7)^2 (k_1 + k_2 - l_6 + l_8)^2 (k_1 + k_2 + l_7 +l_8)^2
}\,. \hskip .5 cm 
\label{GravIntegralE}
\end{eqnarray}

The complete supergravity amplitude is given in
terms of the integrals in \fig{IntegralsThreeLoopFigure}
\begin{eqnarray}
M_4^{(3)} \! & = & \!\Bigl({\kappa \over 2}\Bigr)^8
  \! s_{12} s_{13} s_{14} M_4^\tree  \sum_{S_3}\,
\Bigl[ I^{\rm (a)} +  I^{\rm (b)} + {\textstyle {1\over 2}} I^{\rm (c)} 
  +  {\textstyle {1\over 4}} I^{\rm (d)}\nn \\
&& \hskip 4 cm \null 
 + 2 I^{\rm (e)} + 2 I^{\rm (f)} + 4 I^{\rm (g)} + 
   {\textstyle {1\over 2}}  I^{\rm (h)} 
 + 2 I^{\rm (i)} 
\Bigr] \,.  \hskip .3 cm 
\label{ThreeLoopAmplitude}
\end{eqnarray}
where the numerators of each integral are given in
\tab{NumeratorGravityOldTable}, $\kappa$ is the gravitational
coupling, and $M_4^\tree$ is the supergravity tree amplitude.
In each term, $S_3$ denotes the set of six permutations of three
external legs, say $\{2,3,4\}$, which lead from $(s_{12},s_{23})$
to the six independent ordered pairs of Mandelstam invariants
$s_{12}$, $s_{13}$, $s_{14}$, serving as arguments of the integral.
The numerical coefficients in front of each integral in
\eqn{ThreeLoopAmplitude} are symmetry factors.  They equal 
$4/S$, where $S$ is the number of elements in the discrete
symmetry group of the diagram. 
Due to supersymmetry Ward identities, the 
expression~(\ref{ThreeLoopAmplitude}) is valid for any of
the $256^4$ combinations of four particles from the 256-dimensional
$\NeqEight$ multiplet. 
A remarkable property of the result is that the dimension
$D$ appears explicitly only in the loop integration measure; in
theories with less than maximal supersymmetry we have no
reason to believe that this property will continue to hold.

In this presentation of the amplitude, the UV behavior of the
integrals (e)--(i) is worse than that of the full amplitude.
In particular, the numerator factors for these integrals are quartic
in the loop momenta.  However, as discussed in ref.~\cite{GravityThree},
non-trivial cancellations between diagrams cause the overall 
degree of divergence to be milder, in line with the behavior of
the corresponding amplitude of $\NeqFour$ super-Yang-Mills theory.
In that amplitude, \tab{NumeratorYMTable} shows that each term
is quadratic in the loop momenta.  In the case of a quartic
behavior for $N^{(x)}$, the condition for an integral $I^{(x)}$
in \eqn{IntegralNormalization} to be finite in $D$ dimensions
is $3D + 4 < 20$, or $D<16/3$.  This inequality corresponds to
the $\NeqEight$ finiteness bound~(\ref{OldPowerCount}) proposed in 
ref.~\cite{BDDPR}.  In the case of a quadratic $N^{(x)}$,
the finiteness condition for $I^{(x)}$ is improved to $3D+2 < 20$, 
or $D<6$.

In ref.~\cite{GravityThree} a cancellation between
integrals (e)--(i) in \fig{IntegralsThreeLoopFigure}
was found using \tab{NumeratorGravityOldTable}, and working
in the ``vacuum approximation'' in which external momenta were
neglected.  This approximation was adequate for demonstrating
finiteness for $D<6$, but not for determining the coefficient
of the $D=6$ (potential) divergence.
Below, in \sect{SuperGravityResultsSection}, we will present 
a non-trivial rearrangement of the results in 
\tab{NumeratorGravityOldTable}, so that each $N^{(x)}$
is quadratic in the loop momenta, and hence each contribution satisfies 
the bound~(\ref{SuperYangMillsPowerCount}) with $L=3$.
In \sect{DivergencesSection} we will then integrate each contribution 
near $D=6$, in order to compute the logarithmic divergence.


\section{Color dressing $\boldsymbol{{\cal N}= 4}$ super-Yang-Mills amplitudes}
\label{ColorDressingSection}

Before turning to the case of $\NeqEight$ supergravity, we first
present the complete color-dressed amplitude of $\NeqFour$
super-Yang-Mills theory.  Although the contributing integrals were
presented in ref.~\cite{GravityThree} for use as input into 
the supergravity calculation, they were not explicitly
assembled into the complete color-dressed amplitude.  Here we 
present the explicit color dressing in terms of 
Lie algebra structure constants, $f^{abc}$.
This type of dressing is natural\footnote{One may, 
alternatively, use a color dressing in terms of group generators
in the fundamental representation; such dressings maintain a close
relation with the double-line notation.}
for Feynman diagrams with particles in the adjoint representation. 
It has been used to color-decompose tree and one-loop
amplitudes~\cite{LanceColor}, and to prove the Kleiss-Kuijf
relations between tree-level color-ordered partial
amplitudes~\cite{KleissKuijf}.
It has also played an important role in the recent discovery of 
additional non-trivial tree-level identities~\cite{TreeJacobi}.

Color dressing in terms of structure constant factors is very simple
for {\it non-contact} contributions. Such terms, in which no propagator 
is canceled by a numerator factor, are those in the numerator
factors $N^{(x)}$ in \tab{NumeratorYMTable} that do not contain a 
factor of $l_i^2$.
(The only explicit factor of $l_i^2$ appears in $N^{{\rm (i)}}$; 
there are also implicit factors of $l_i^2$ in expressions such as
$\tau_{26} = s_{26} - l_6^2$ in integral (h).)
The non-contact contributions are detectable from the maximal cuts,
which have only three-point amplitudes.  In $\NeqFour$ super-Yang-Mills
theory, all such amplitudes are proportional to a factor of $f^{abc}$.
Thus the appropriate color factor for a non-contact term in a 
numerator $N^{(x)}$ is found simply by dressing each three-point
vertex in the corresponding parent graph $(x)$ in 
\fig{IntegralsThreeLoopFigure} with an $f^{abc}$.  

The color factors for {\it contact-term} contributions with a canceled
propagator are somewhat less obvious, because they are determined from
cuts containing four- or higher-point amplitudes.  Such amplitudes are
not proportional to a single product of $f^{abc}$'s, but contain
multiple terms.  The color factors can always be expressed as sums of
products of $f^{abc}$'s, but it is conceivable that not all such
products would be of the same form as one of the (nonvanishing)
non-contact terms.  However, for the three-loop four-point amplitude,
we find that they are all of the same form, so that the contact terms
in each $N^{(x)}$ can be consistently dressed with the same graphical
color factor as the non-contact terms.  We have confirmed the consistency
of this assignment by evaluating all the generalized cuts
in~\fig{RealCutsFigure} using tree amplitudes dressed with full color
factors as the building blocks.  The evaluation of these cuts is a
complete check of the color dressing. The check requires use of color
Jacobi identity rearrangements of the type described in
ref.~\cite{LanceColor}.  For general amplitudes, we expect the same
property to hold:  Once the contact terms are assigned to parent integrals 
in a way consistent with unitarity cuts that 
use color-{\it ordered} amplitudes, 
then the full color-{\it dressed} amplitudes should be obtained
simply by dressing the parent integrals with an $f^{abc}$ at each
three vertex.

In summary, the fully color-dressed three-loop four-point $\NeqFour$
super-Yang-Mills amplitude is given by,
\begin{eqnarray}
{\cal A}_4^{(3)} \! & = & \! -
{1\over 4}\,g^8\, s_{12} s_{14} A_4^\tree \sum_{S_4}\, 
\Bigl[ C^{\rm (a)} I^{\rm (a)} 
  + C^{\rm (b)} I^{\rm (b)} 
  + {\textstyle {1\over 2}} C^{\rm (c)} I^{\rm (c)} 
  +  {\textstyle {1\over 4}} C^{\rm (d)} I^{\rm (d)}\nn \\
&& \null \hskip 2 cm 
 + 2 C^{\rm (e)} I^{\rm (e)} 
 + 2 C^{\rm (f)} I^{\rm (f)} + 4 C^{\rm (g)} I^{\rm (g)} + 
   {\textstyle {1\over 2}}  C^{\rm (h)} I^{\rm (h)} 
 + 2 C^{\rm (i)} I^{\rm (i)} 
\Bigr] \,,  \hskip .3 cm 
\label{ThreeLoopYMAmplitude}
\end{eqnarray}
where $g$ is the gauge coupling, $C^{(x)}$ are the color
factors, and $I^{(x)}(s,t)$ are $D$-dimensional loop integrals
corresponding to the nine diagrams in \fig{IntegralsThreeLoopFigure}.
The sum runs over the 24 independent permutations of legs
$\{1,2,3,4\}$, denoted by $S_4$.  The permutations in $S_4$
act on both kinematic and color labels.  In the case of the
gravity amplitude~(\ref{ThreeLoopAmplitude}) the $S_4$ sum could
be collapsed to an $S_3$ sum, holding leg 1 fixed,
because the summand exhibits an additional symmetry due to the 
kinematic identities $s_{12}=s_{34}$, $s_{13}=s_{24}$, and
$s_{14}=s_{23}$.
In the Yang-Mills case, the presence of color factors with no such
manifest symmetry prevents us from collapsing the sum.
Other than this minor difference, note the similarity of the gravity
(\ref{ThreeLoopAmplitude}) and gauge theory amplitudes
(\ref{ThreeLoopYMAmplitude}), including the symmetry factors of each
integral.  In \eqn{ThreeLoopYMAmplitude}, $A_4^\tree$ is the
color-ordered tree amplitude $A_4^\tree(1,2,3,4)$.  With the factor of
$s_{12} s_{14}$, it has the required overall crossing and Bose
symmetry of color-dressed gauge theory amplitudes.

\begin{figure}[t]
\centerline{\epsfxsize 1.7 truein \epsfbox{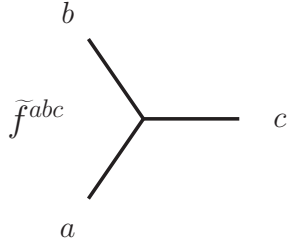}}
\caption[a]{\small Color factors for the super-Yang-Mills amplitude are
obtained simply by dressing
the diagrams in \fig{IntegralsThreeLoopFigure} with an $\f^{abc}$ at each
three vertex.  All contact terms pick up the same color factor as the parent
(non-contact) diagrams with which they are associated.}
\label{ColorDressFigure}
\end{figure}

As illustrated in \fig{ColorDressFigure}, the color factors
$C^{(x)}$ are given by dressing each three-vertex of the parent
diagrams in \fig{IntegralsThreeLoopFigure} with modified structure
constants,
\begin{equation}
\f^{abc} = i \sqrt{2} f^{abc} = \Tr([T^a, T^b] T^c)\,,
\label{fabcdefn}
\end{equation}
where $f^{abc}$ are the standard structure constants, and the
hermitian generators are normalized via
$\Tr[T^a T^b] = \delta^{ab}$. The $\f^{abc}$ should follow the 
clockwise ordering of the parent diagram vertices, 
respecting the ordering of each vertex in
\fig{IntegralsThreeLoopFigure}.  The color factor associated with 
each integral is then easy to write down.  For example, 
for diagrams (a) and (i) we have,
\begin{eqnarray}
&& C^{\rm (a)} = \g{1}{5}{6} \g{2}{9}{5} \g{3}{13}{14} \g{4}{11}{13}
\g{6}{7}{8} \g{7}{9}{12} \g{8}{10}{11} \g{10}{12}{14} \,, \nn \\
&& C^{\rm (i)} = \g{1}{8}{5} \g{2}{10}{7} \g{3}{6}{9} \g{4}{13}{12}
\g{5}{11}{13} \g{6}{12}{14} \g{7}{14}{11} \g{8}{10}{9} \,.
\end{eqnarray}
The other factors work similarly.
The factors for the different permutations in \eqn{ThreeLoopYMAmplitude}
are obtained by permuting the external labels $\{1,2,3,4 \}$.


\section{$\boldsymbol{{\cal N}=8}$ supergravity amplitude with 
manifest ultraviolet behavior}
\label{SuperGravityResultsSection}

\begin{figure}[t]
\centerline{\epsfxsize 5.2 truein \epsfbox{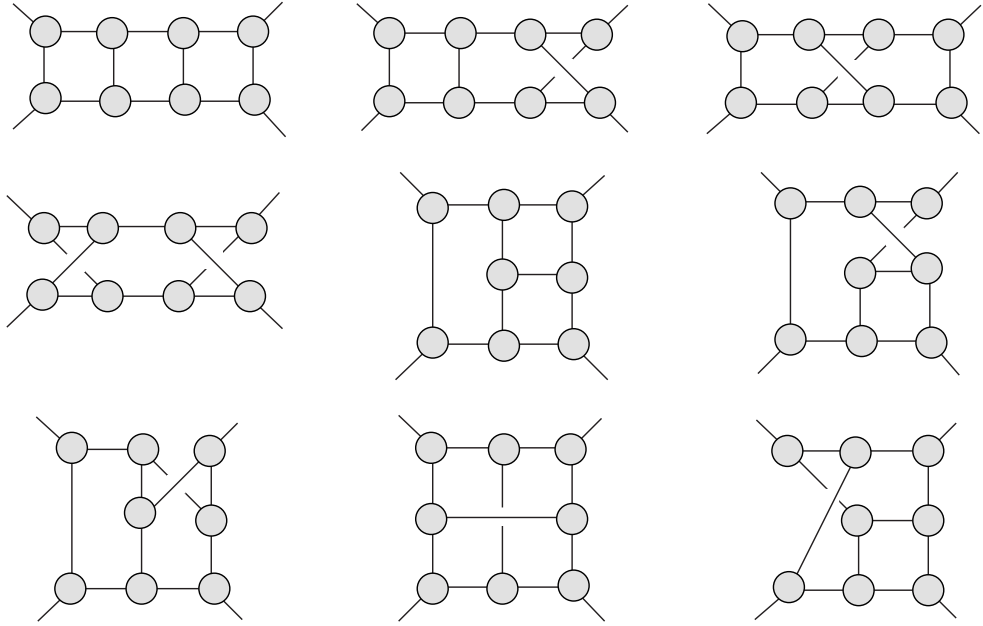}}
\caption[a]{\small The nine maximal cuts used to determine the
integrands, up to contact terms.  All propagator lines are taken to be
cut, with the momenta satisfying on-shell conditions. The vertices
represent on-shell three-point amplitudes.}
\label{MaximalCutsFigure}
\end{figure}

The three-loop four-point $\NeqEight$ amplitude presented in
ref.~\cite{GravityThree} has better UV behavior than each integral
taken separately.
It is therefore natural to try to find a representation in which each
integral exhibits the behavior of the complete amplitude.  One may
attempt to manipulate the numerator factors in
\tab{NumeratorGravityOldTable}, by moving terms between different
integral contributions until the leading two powers of loop momenta
cancel.  This procedure turns out to be significantly more difficult
than simply reconstructing the amplitude from scratch using the method
of maximal cuts~\cite{FiveLoop}, as we do here.  By isolating in any
one cut a small number of terms in the amplitude, it becomes much
simpler to find ans\"{a}tze for new compact forms of the amplitude, and to
arrange the amplitude so that no term has a worse behavior than the
complete amplitude.

We start with an ansatz for the amplitude in terms of Feynman
integrals with numerator polynomials containing arbitrary parameters.
We require that each numerator polynomial is at most
quadratic in the loop momenta,
\begin{equation}
N^{(x)} = \sum a_{ij}^{(x)}\, l_i\cdot l_j +  
  \sum b_{i,j,m,n}^{(x)} \, l_i\cdot k_j \, l_m \cdot k_n 
 + \sum c_{i,j}^{(x)}\, l_i\cdot k_j 
+ d^{(x)} \,,
\label{NumeratorAnsatz}
\end{equation}
where $a_{ij}^{(x)}$, $b_{i,j,m,n}^{(x)}$, $c_{i,j}^{(x)}$ 
and $d^{(x)}$ are polynomials in
the external momenta containing free parameters.  In the sums we
include only those terms not simply related to the others via momenta
conservation.  To determine the parameters we replace the numerators
in \eqn{ThreeLoopAmplitude} with the numerators
(\ref{NumeratorAnsatz}), and compare the cut of the ansatz against the
cut of the amplitude,
\begin{equation}
\sum_{\rm states} A^\tree_{(1)} A^\tree_{(2)} A^\tree_{(3)} \cdots 
A^\tree_{(m)} \,,
\end{equation}
using kinematics that place all cut lines on-shell, $l_i^2 = 0$.
Although the comparison can be done analytically, it is generally
simplest to generate kinematic solutions
numerically~\cite{FiveLoop}. If no solution to the cut conditions are
found, then we enlarge the ansatz until one is
found. For the $\NeqEight$ supergravity three-loop four-point
amplitude, the quadratic numerator
ansatz~(\ref{NumeratorAnsatz}) is sufficient.

We start by analyzing cuts with the maximum number of cut propagators.
The nine distinct---up to relabellings of external legs---maximal cuts
for the three-loop four-point amplitude are displayed in
\fig{MaximalCutsFigure}.  As discussed in ref.~\cite{FiveLoop}, by
choosing appropriate $D=4$ kinematics, we can force all cut lines to
be gluons in the super-Yang-Mills case.  Similarly, in the
supergravity case, the same kinematics will force all cut lines to be
gravitons.  Remarkably, these ``singlet cuts'' turn out to be sufficient
to determine the non-contact terms in four point amplitudes in
$\NeqFour$ super-Yang-Mills theory and in $\NeqEight$ supergravity,
in all known cases.

\begin{figure}[t]
\centerline{\epsfxsize 4.7 truein \epsfbox{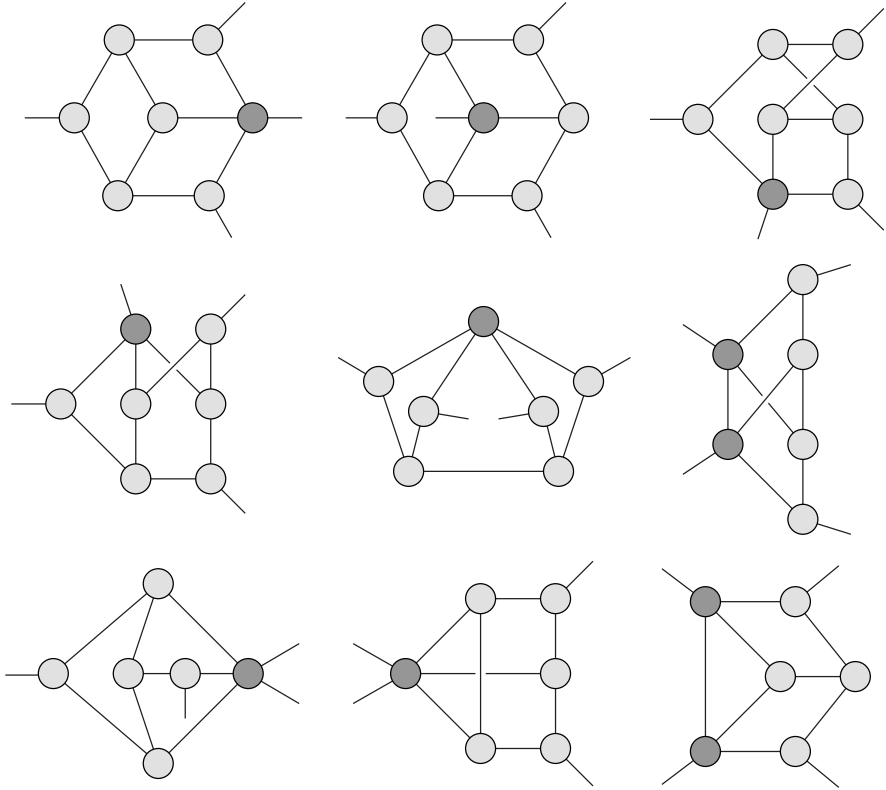}}
\caption[a]{\small Nine near-maximal cuts that are especially helpful for
obtaining the contact terms for the new form of the supergravity amplitude.  All
propagator lines are taken to be cut, with the momenta satisfying
on-shell conditions. The darker vertices represent on-shell four-
and five-point amplitudes; the lighter ones depict three-point amplitudes.}
\label{ComplexCutsFigure}
\end{figure}

Contact terms, containing  numerator factors of $l_i^2$, cannot be
determined from maximal cuts, because the $l_i^2$ are all set to zero
in these kinematics.
To determine the contact terms, we systematically reduce the number of
cut lines, until all potential contact terms are identified.  In
\fig{ComplexCutsFigure}, we show some near-maximal cuts that are
particularly helpful for determining all the contact terms in the
three-loop four-point amplitudes.  In this figure every line
represents a cut propagator ({\it i.e.}, with the momenta taken on
shell).  In these cases, again by appropriate choices of
kinematics~\cite{FiveLoop}, we can force almost all lines to be
gluons or gravitons.

By carrying out this procedure in $D$ dimensions, we can apply the
maximal cut method to any theory, including QCD.  However, to take
advantage of four-dimensional spinor simplifications and singlet cuts,
we restrict the momenta to be four-dimensional, which potentially can
drop contributions.  For maximally supersymmetric four-point
amplitudes, it turns out that the four-dimensional
cuts appear to be sufficient for determining all contributions.  (As
mentioned in the introduction, in non-maximally supersymmetric
theories, and in maximally supersymmetric amplitudes with more than
four external states, amplitudes typically contain extra terms
proportional to the $(-2\eps)$ dimensional components of the loop
momenta.)  We can also take advantage of a variety of pictorial
rules~\cite{BRY,FiveLoop,FreddyMaximal,TreeJacobi} for obtaining more
complicated contributions from simpler ones.

To find compact forms of the $\NeqEight$ amplitude we evaluated 98
possible maximal and near-maximal cut topologies that contain up to
two contact terms.  The nine maximal cuts are depicted in 
\fig{MaximalCutsFigure}.   The set of maximal cuts begins with all graphs
with only cubic vertices.  For the applications to four-point
amplitudes in maximally supersymmetric theories, 
we can discard the subset of graphs which contain triangle subgraphs 
or those which are one-particle reducible.  We can also discard
graphs containing a two-particle cut that exposes such graphs at
two loops.  Such cuts will lead to vanishing contributions.
The surviving nine maximal cuts are in one-to-one correspondence
with the nine parent integrals depicted in~\fig{IntegralsThreeLoopFigure}.
Then 27 distinct single contact term diagrams are obtained
from the nine maximal cut topologies by systematically collapsing one
propagator, removing ones related by
symmetry. Similarly  62 double contact terms are obtained by
collapsing two propagators, again removing ones related by symmetry.
The 18 cuts illustrated in \figs{MaximalCutsFigure}{ComplexCutsFigure}
turn out to be sufficient for finding a representation of the
amplitude which exhibits quadratic dependence on the loop momenta in
the numerators, and thus manifestly obeys the
bound~(\ref{SuperYangMillsPowerCount}) for $L=3$.  Of course, these
cuts by themselves do not rule out other potential contributions,
including those with three or more collapsed propagators.  For this we
rely on the generalized cuts in \fig{RealCutsFigure}, evaluated in $D$
dimensions.

\begin{figure}[t]
\centerline{\epsfxsize 4.7 truein \epsfbox{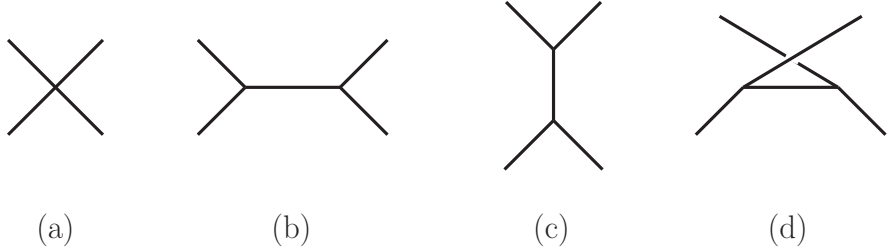}}
\caption[a]{\small A contact term can be assigned to 
parent diagrams by inserting a propagator multiplied by a
numerator factor which cancels it. A contact term (a) can be
distributed amongst the three propagator diagrams (b), (c) and (d), by
including appropriate numerator factors to cancel the propagators.}
\label{ContactFigure}
\end{figure}

In general there is a large freedom in assigning numerator factors
containing loop momenta to parent integrals, especially when no color
factors are present to distinguish different integrals, as in the
supergravity case.  For example, scalar products of loop momenta with
external momenta can often be rearranged using the observation that
such products can be expressed as differences of inverse propagators,
{\it e.g.}, $2 k_i \cdot l_j = (k_i+l_j)^2 - l_j^2$.  As another
particularly simple example, note that when assigning contact terms to
parent diagrams, as illustrated in \fig{ContactFigure}, any four-point
contact term (a) can be expressed as a combination of propagator
diagrams (b)--(d) with appropriate numerator factors.  The precise
form is subject only to the constraint that the contributions add up
to the contact term.  Moreover, by using momentum conservation
(and on-shell conditions on any cut legs) we can move contributions
around within a larger diagram.

\begin{table*}
\caption{
A form for the $\NeqEight$ supergravity numerator factors $N^{(x)}$ for 
the integrals $I^{(x)}$ in \fig{IntegralsThreeLoopFigure}
appearing in \eqn{ThreeLoopAmplitude}, in which all terms
are at most quadratic in the loop momenta.
The first column labels the integral, the second column the relative
numerator factor. 
\label{NumeratorGravityTable} }
\vskip .4 cm
\begin{tabular}{||c|c||}
\hline
Integral $I^{(x)}$ & $N^{(x)}$ for $\NeqEight$ Supergravity  \\
\hline
\hline
(a)--(d) 
& $
\vphantom{\Bigr|}
 [s_{12}^2]^2
$ \\
\hline
(e)--(g) & $\vphantom{\Bigr|}
 s_{12}^2 \, \tau_{35} \, \tau_{46}
$  \\
\hline
(h) & $\;
(s_{12} (\tau_{26} + \tau_{36}) + s_{14}(\tau_{15}+\tau_{25})+s_{12} s_{14})^2
\;$ \\
& $\; \null
   + (s_{12}^2 (\tau_{26} + \tau_{36})
   -  s_{14}^2 (\tau_{15} + \tau_{25})) 
             (\tau_{17} + \tau_{28} + \tau_{39} + \tau_{4,10})
         \; $  \\
    &  $ \; \null
      + s_{12}^2 (\tau_{17}\, \tau_{28} + \tau_{39}\, \tau_{4,10})
      + s_{14}^2 (\tau_{28}\, \tau_{39} + \tau_{17}\, \tau_{4,10} )
           + s_{13}^2  (\tau_{17}\, \tau_{39} + \tau_{28} \, \tau_{4,10}) 
\;     \vphantom{\Bigl( \Bigr)_{A_A} }
         $   \\
\hline
(i) & $
\vphantom{\bigl|_{A_A}} 
(s_{12}\, \tau_{45} - s_{14} \, \tau_{46})^2 
- \tau_{27} (s_{12}^2 \,\tau_{45} + s_{14}^2 \,\tau_{46})
- \tau_{15} (s_{12}^2 \,\tau_{47} + s_{13}^2 \,\tau_{46})
 \; $ \\
& $ \; \null
- \tau_{36} (s_{14}^2\, \tau_{47} + s_{13}^2 \,\tau_{45})
+ l_5^2 \, s_{12}^2 \,s_{14} 
+ l_6^2 \, s_{12} \, s_{14}^2 
- {1\over 3} l_7^2 \, s_{12} \, s_{13} \, s_{14}
\; $ \\
\hline
\end{tabular}
\end{table*}

A particularly compact form for
the supergravity amplitude is shown in \tab{NumeratorGravityTable}.
The complete amplitude is given by \eqn{ThreeLoopAmplitude}, 
in which the numerators $N^{(x)}$ in \tab{NumeratorGravityOldTable}
are replaced by those in \tab{NumeratorGravityTable}.
Each numerator respects the symmetry of its corresponding diagram.
Because the new form matches the $D$-dimensional cuts displayed
in \fig{RealCutsFigure}, it is completely equivalent to the earlier
form.  The two forms are, however, rather non-trivially related.
In particular, the form in \tab{NumeratorGravityOldTable} has
numerator terms which are quartic in loop momenta, while the form
in \fig{NumeratorGravityTable} is merely quadratic.
The form in \tab{NumeratorGravityTable} is by no means unique. 
The large freedom, mentioned previously, in assigning
numerator factors to a parent graph implies that there are
continuous families of numerators with the same quadratic behavior
that satisfy all cut conditions.  As a check, we have evaluated the
logarithmic divergence of the various forms at $D=6$, and we find that
it is independent of each of the free parameters, as expected.

\section{Divergences in higher dimensions}
\label{DivergencesSection}

As demonstrated in ref.~\cite{BDDPR}, the two-loop four-graviton 
$\NeqEight$ supergravity  amplitude saturates the finiteness
bound (\ref{SuperYangMillsPowerCount}), having a critical dimension $D_c=7$.
The values of the two-loop divergence in dimensions
between $D=7$ and $D=11$ were also calculated in that reference,
in dimensional regularization.  In
this section we carry out a similar analysis at three loops, 
using the new representations of the amplitudes given in
\tab{NumeratorGravityTable}.  We prove that in $\NeqEight$
supergravity the bound (\ref{SuperYangMillsPowerCount}) is saturated
at $L =3$; that is, a logarithmic UV divergence is present 
in $D=6$.  We also give the explicit values of the dimensionally-regulated
power-law divergences in $D=7, 9, 11$.  Other than the
critical dimension $D_c=6$, odd dimensions are also interesting
because of the connection to M theory dualities~\cite{GreenVanhove} in
dimensions $D=9,11$.  From a technical standpoint, it is easier to
compute the three-loop divergences in odd dimensions than it is in even
dimensions.  In odd dimensions the only divergences that arise in
dimensional regularization are from two-loop subdivergences.

\subsection{Divergence in the critical dimension $D=6$.}

We already know that for $\NeqFour$ super-Yang-Mills theory the
finiteness bound is saturated through at least four
loops~\cite{Finite}.  Is the bound (\ref{SuperYangMillsPowerCount})
saturated for $\NeqEight$ supergravity as well?  It would of course be
rather surprising if $\NeqEight$ supergravity were better behaved
in the UV than $\NeqFour$ super-Yang-Mills theory ---
but surprises have happened before.  It is
nevertheless important to confirm that no further ``hidden''
cancellations exist.
Because we now have a compact analytic expression for the
amplitude, without additional spurious divergences that cancel between
integrals, we can settle the issue simply by evaluating the integrals
near $D=6$ and assembling the amplitude in this dimension. 
If the amplitude diverges in $D=6$, then no further cancellations
exist at three loops, beyond those found in ref.~\cite{GravityThree},
and exhibited manifestly by the numerator factors in 
\tab{NumeratorGravityTable}.

In its critical dimension, where an integral first develops a
divergence, the divergence is logarithmic.  This implies that the
residue of the $1/\epsilon$ pole is constant, after all factors of
external momentum in the numerator are taken out of the integral.  To
evaluate this constant we may choose the external momenta and
invariants in any convenient way.  In particular, we may expand the
integrand for small external momenta and keep the leading term.  A key
advantage of the new representation of the amplitude in
\tab{NumeratorGravityTable}, compared to the original one in
\tab{NumeratorGravityOldTable}, is that no integral has a critical
dimension below the expected critical dimension of the amplitude. 
This simplifies the evaluation of the divergence because it receives
contributions only from the leading term in the small-momentum expansion. 

In addition we can drop all integrals with a critical dimension larger
than six.  Inspecting \tab{NumeratorGravityTable},
\fig{IntegralsThreeLoopFigure}, and \eqn{IntegralNormalization},
we see that integrals (a)--(d) are all
finite in $D=6$, so we need only evaluate the integrals (e)--(i).
These integrals have numerators quadratic in the loop momenta. We
expand at small external momenta, keeping only the UV-divergent 
integrals which are independent of the external momenta,
apart from the explicit factors of external momenta appearing 
in the numerator.  Lorentz invariance implies that we can replace
\begin{equation}
\int d^D q_1 d^D q_2 d^D q_3 \; q_i^\mu q_j^\nu f(q_1, q_2, q_3) \rightarrow
{\eta^{\mu\nu} \over D}
\int d^D q_1 d^D q_2 d^D q_3 \; q_i\cdot q_j f(q_1, q_2, q_3) \,,
\label{LorentzSubs}
\end{equation}
because $\eta_{\mu\nu}$ is the only available tensor 
that can have a divergent coefficient.
Contracting both sides with $\eta_{\mu\nu}$ shows that the
prefactor on the right-hand side is indeed $1/D$. For the purpose of
computing the leading UV singularity in the critical dimension, this
identity allows us to replace any two-tensor integral with simpler
scalar integrals, obtained by rewriting $q_i\cdot q_j$ 
as a linear combination of inverse propagators.

\begin{figure}[t]
\centerline{\epsfxsize 4.5 truein \epsfbox{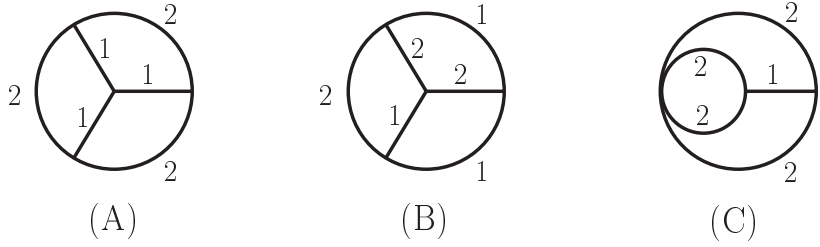}}
\caption[a]{\small Vacuum-like diagrams describing the leading behavior.
The labels 1,2 indicate whether the corresponding propagator appears 
to the first or second power.}
\label{VacuumFigure}
\end{figure}

Keeping only the leading terms at small external momenta and using
\eqn{LorentzSubs}, we replace the diagrams in
\fig{IntegralsThreeLoopFigure} with the simpler vacuum diagrams shown in
\fig{VacuumFigure}.  We have,
\begin{eqnarray}
&&
I^{\rm (a)} \rightarrow 0\,, \hskip .4cm 
I^{\rm (b)} \rightarrow 0\,, \hskip .4cm 
I^{\rm (c)} \rightarrow 0\,, \hskip .4cm 
I^{\rm (d)} \rightarrow 0\,, \hskip .4cm  \nn\\
&&
I^{\rm (e)} \rightarrow {-s_{12}^3 \over 3} V^{\rm (A)}\,, \hskip .4 cm
I^{\rm (f)} \rightarrow {-s_{12}^3 \over 3} V^{\rm (B)}\,, \hskip .4 cm 
I^{\rm (g)} \rightarrow {-s_{12}^3 \over 3} V^{\rm (B)}\,, \hskip .4 cm  \nn\\
&&
I^{\rm (h)} \rightarrow {2 \over 3}(2s_{13}^3-3s_{12}s_{13} s_{14}) 
   V^{\rm (B)}
 +(-s_{13}^3+ s_{12} s_{13} s_{14}) V^{\rm (C)} \,, \hskip .4 cm \nn\\
&& 
I^{\rm (i)} \rightarrow {1 \over 6}(2s_{13}^3 - 5 s_{12} s_{13} s_{14})
  V^{\rm (A)}-{1 \over 3}(s_{13}^3+3s_{12} s_{13} s_{14}) V^{\rm (B)}\,, 
 \hskip .4 cm 
\label{VacSubs}
\end{eqnarray}
where we used 
\begin{equation}
s_{12}^3 + s_{13}^3 + s_{14}^3 = 3 s_{12} s_{13} s_{14}
\label{scubicid}
\end{equation}
to simplify the expressions, and we set $D=6$ in the coefficients. 
Multiplying the above
expressions by their numerical prefactors in \eqn{ThreeLoopAmplitude}
and summing over the permutations, it is not hard to find that the leading UV
divergence in $D=6$ is
\begin{eqnarray}
M_4^{(3), D=6-2\ep} \Bigr|_{\rm pole}\! & = & \!-\Bigl({\kappa \over 2}\Bigr)^8
  \! (s_{12} s_{13} s_{14})^2 M_4^\tree  
\Bigl[  10(V^{\rm (A)} + 3V^{\rm (B)}) \Bigr] \,.  \hskip .3 cm 
\label{VacSubsFinal}
\end{eqnarray}
The coefficient of the vacuum-like
diagram, $V^{\rm (C)}$, which arises only from $I^{\rm (h)}$,
vanishes after the permutation sum. These features are 
a consequence of the specific representation of
the numerators given in \tab{NumeratorGravityTable}.

\begin{figure}[t]
\centerline{\epsfxsize 4.4 truein \epsfbox{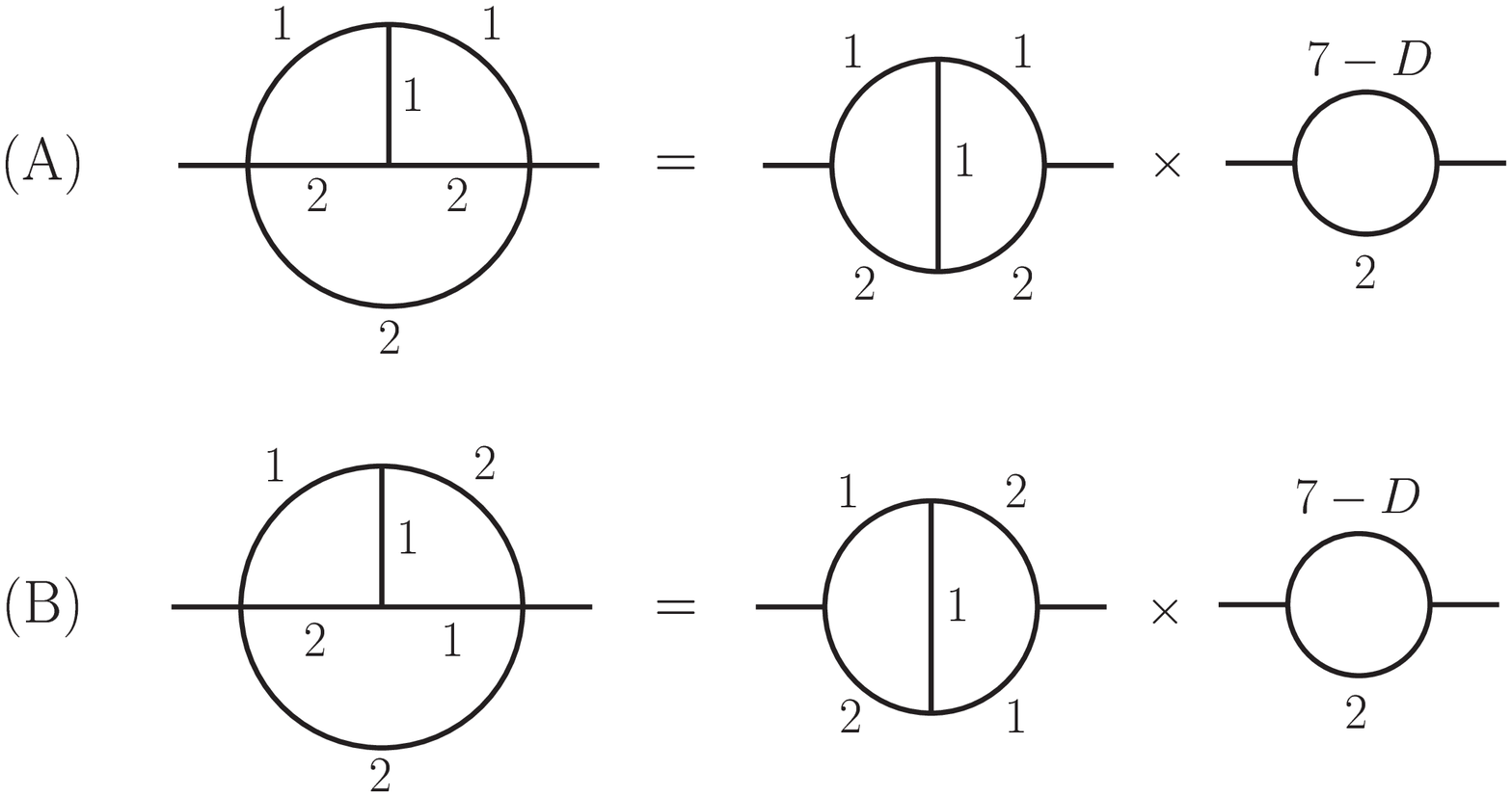}}
\caption[a]{\small Reduction of three-loop propagator diagrams
appearing in the $D=6-2\e$ UV divergences of $\NeqEight$ supergravity.
The numbers near each propagator indicates the power or index to which it
is raised.
}
\label{UVpolesFigure}
\end{figure}

To determine whether the leading $1/\eps$ pole in \eqn{ThreeLoopAmplitude}
cancels, we must evaluate the UV behavior of the 
vacuum integrals (A) and (B) in \fig{VacuumFigure}.
A convenient means for doing so is to restore some momentum dependence
in order to regulate the infrared divergences.  We can 
re-interpret the calculation as that of a typical propagator integral,
simply by injecting some arbitrary external momentum at 
appropriately-chosen vertices, and applying standard 
techniques developed for this problem.  (See for example the
review by Grozin~\cite{GrozinLectures}.)  
Because the divergence is logarithmic in the critical dimension, 
the injected momentum does not appear in the leading UV pole in
dimensional regularization; thus it has no effect on the result.
We therefore promote the vacuum-like diagrams (A) and (B) in
\fig{VacuumFigure} to propagator diagrams, while leaving all internal
lines massless, as shown in \fig{UVpolesFigure}.  Then we can use
dimensional analysis to simplify each three-loop propagator diagram
down to a product of a two-loop propagator diagram and a one-loop
bubble diagram.  (Although we do not need it here, diagram (C) in
\fig{VacuumFigure} can be evaluated in a similar fashion.)

As shown in \fig{UVpolesFigure}, to evaluate the integrals, we factor out
a two-loop propagator diagram formed from the two upper loops 
on the left side of diagrams (A) and (B). By dimensional analysis, the two-loop
propagator subintegrals are given by
\begin{equation}
P^{\rm (A)}_\twoloop = { K^{\rm (A)}_\twoloop \over (4\pi)^D }
 {1\over (l^2)^{7-D}} \,, \hskip 2 cm 
P^{\rm (B)}_\twoloop = { K^{\rm (B)}_\twoloop \over (4\pi)^D }
 {1\over (l^2)^{7-D}} \,,
\end{equation}
where $l$ is the momentum flowing through the two-loop propagator
diagram and $K^{\rm (A)}_\twoloop$ and $K^{\rm (B)}_\twoloop$ are constants to
be determined by explicit integration. The factor of $1/(l^2)^{7-D}$
effectively results in a remaining one-loop bubble integral with a
propagator raised to the power $(7-D)$.  The two-loop propagator
integrals in each case are finite in $D=6$.  We evaluated them using
integration by parts~\cite{IBP}, with the result,
\begin{equation}
K^{\rm (A)}_\twoloop = 1 \,, \hskip 2 cm 
K^{\rm (B)}_\twoloop = \zeta_3 - {1\over3} \,.
\label{Kvalues}
\end{equation}

The remaining one-loop bubble integrals are also straightforward to
evaluate; for arbitrary exponents $n_1$ and $n_2$ of the two 
propagators, they are given by~\cite{GrozinLectures}
\begin{equation}
I^{\rm bubble}(n_1,n_2)\ \equiv -i\int\frac{d^Dp}{(2\pi)^D}
\frac{1}{\bigl(\,(p+k)^2\,\bigr)^{n_1} (p^2)^{n_2} }
\ =\ \frac{(-1)^{n_1+n_2}}{(4\pi)^{D/2}}
G(n_1,n_2) (-k^2)^{-(n_1+n_2-D/2)}\,,
\label{Bubble}
\end{equation}
where
\begin{equation}
G(n_1,n_2)\ =\ { \Gamma(-D/2+n_1+n_2) \Gamma(D/2-n_1)\Gamma(D/2-n_2)
                \over \Gamma(n_1) \Gamma(n_2) \Gamma(D-n_1-n_2) } \,.
\label{DefineG}
\end{equation}

In $D=6-2\eps$, for the two cases in \fig{UVpolesFigure} we have $n_1
= 1+2\eps$ and $n_2 = 2$.  The bubble integral in both cases provides
the UV divergence in $D=6-2\e$,
\begin{equation}
G(1+2\e,2)\ =\ { \Gamma(3\e) \Gamma(2-3\e)\Gamma(1-\e)
        \over \Gamma(1+2\e) \Gamma(2) \Gamma(3-4\e) }
\ =\ { 1 \over 6 \e }\ +\ \Ord(1) \,.
\label{UVdivbubbleb}
\end{equation}
Recalling the normalization of the integrals $I^{(x)}$ defined
in \eqn{IntegralNormalization}, and collecting factors from
\eqns{Kvalues}{Bubble}, we obtain the UV singularity
of the vacuum-like diagrams (A) and (B),
\begin{eqnarray}
V^{\rm (A)} &=& -\frac{1}{(4\pi)^9}\left[{1\over6 \e}\ +\ \Ord(1)\right]
\,, \label{I112221} \\
V^{\rm (B)} &=& -\frac{1}{(4\pi)^9}
\left[{1\over6 \e} \biggl( \zeta_3 - {1\over3} \biggr)\ +\ \Ord(1)\right]\,.
\label{Results}
\end{eqnarray}
We have confirmed these results by direct numerical integration 
of the vacuum-like diagrams, using a mass regulator to define the integrals.
Using \eqn{VacSubsFinal}, we obtain
\begin{eqnarray}
M_4^{(3), D=6-2\ep} \Bigr|_{\rm pole}\! & =
 &\frac{1}{\epsilon}\,{5\zeta_3\over (4\pi)^9} \!
\Bigl({\kappa \over 2}\Bigr)^8
 \! (s_{12} s_{13} s_{14})^2 M_4^\tree  \,.  \hskip .3 cm 
\label{ThreeLoopD6Div}
\end{eqnarray}
Note that the simple functional dependence of \eqn{ThreeLoopD6Div} on the
kinematic variables $s_{12}$, $s_{13}$ and $s_{14}$ is
fixed by dimensional analysis and Bose symmetry, given that it should
contain a factor of $s_{12} s_{13} s_{14} M_4^\tree$.
The identity~(\ref{scubicid}) can be used to help establish this fact.

Notice also the cancellation of the rational parts of
$K_\twoloop^{\rm (A)}$ and $K_\twoloop^{\rm (B)}$,
in the combination $V^{\rm (A)} + 3 V^{\rm (B)}$ appearing in
\eqn{VacSubsFinal}.  Thanks to this cancellation, the logarithmic UV
divergence~(\ref{ThreeLoopD6Div}) possesses a uniform degree of
transcendentality (in which $\zeta_n$ is assigned degree $n$,
and rational numbers degree zero).  This property is common to
infrared-regulated amplitudes, near $D=4$, in $\NeqFour$ 
super-Yang-Mills theory and $\NeqEight$ supergravity;
here we see it persists to the level of the three-loop UV singularity
at $D=6$.  

As mentioned earlier, the fact that \eqn{ThreeLoopD6Div}
is nonzero establishes that, in comparison with $\NeqFour$
super-Yang-Mills theory, $\NeqEight$ supergravity is no better
behaved (as well as no worse behaved) in the ultraviolet
through three loops.

\subsection{UV divergences in odd dimensions}

Next we turn to the computation of three-loop divergences
in dimensions above six.  To simplify the analysis, we restrict our
attention to the behavior near odd values of $D$, namely $D=7,9,11$.

According to the convergence theorem, a Feynman integral is convergent
if the degree of divergence of all one-particle-irreducible
subintegrals is negative.
Together with the BPHZ subtraction it also implies that, after
subtracting all subdivergences, the remaining overall divergence
(arising when all loop momenta are scaled to infinity at the same
rate) has a local (polynomial) dependence on external momentum
invariants.
In dimensional regularization it is easy to see that, in odd
dimensions and at an odd loop order, no such overall divergences may
exist. Indeed, the dimension of any such loop integral is odd, and so the
result of the integration depends on a half-integer power of momentum
invariants.  However, such a dependence would be nonlocal.  Hence its
coefficient must be finite after the subtraction of all subdivergences.
Thus, after subtracting the subdivergences, all three-loop integrals
must be finite in odd dimensions. For the same reason, no subtraction
is necessary for one-loop subintegrals in odd dimensions.  
Therefore, the first divergent
subintegrals have two loops, and they completely determine the leading
UV divergence of the three-loop integrals to which they belong.

\begin{figure}[t]
\centerline{\epsfxsize 4.7 truein \epsfbox{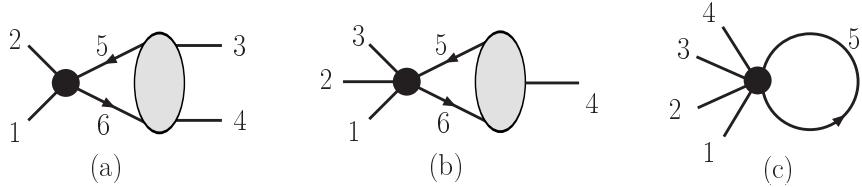}}
\caption[a]{\small The two-loop subdivergence contributions to the
  three-loop local divergence. The black vertex represents a two-loop
  local divergence.  In dimensional regularization, near a dimension
  greater than four, contributions (b) and (c)
  vanish because they are proportional to a positive power of a
  vanishing invariant. The complete contribution is
  given by summing over the inequivalent permutations of external
  legs.}
\label{AmplDivergenceFigure}
\end{figure}

The above arguments hold for general off-shell integrals. Our case is
somewhat simpler due to the masslessness of the external momenta.
Consider the potential two-loop subdivergences shown in 
\fig{AmplDivergenceFigure}. In principle, there can be contributions
from two-loop four-, five- and six-point subdivergences.  (There is no
three-point subdivergence in any supergravity theory.)  However, from
dimensional analysis in $D>4$ dimensions we can easily show that both
diagrams (b) and (c) vanish, because they are
proportional to vanishing invariants. For example,
inserting a five-point two-loop divergence into the third loop, 
as shown in figure \ref{AmplDivergenceFigure}(b), amounts to computing
\begin{eqnarray}
\int \frac{d^Dq}{(2\pi)^D}\frac{P(q,k_1,k_2)}{q^2\;(q+k_4)^2} \sim 
(k_4^{2})^{D-4 +n_q/2} \,.
\end{eqnarray}
Here $P(q,k_1,k_2)$ is a polynomial depending on the loop and
external momenta, and $n_q$ counts the power of $q$ in the
numerator of a given term.  Even for terms with no powers of $q$ in
the numerator, for $D\ge6$ the integral is proportional to a positive
power of $k_4^2 = 0$ and therefore vanishes.  With additional powers
of loop momenta $q$ in the numerator, the vanishing is even stronger.

Similarly, integral (c) in \fig{AmplDivergenceFigure} must vanish,
because no external invariant appears.  It is interesting to note that
the three-loop integrals (d), (h) and (i) in
\fig{IntegralsThreeLoopFigure} do not have two-loop four-point
subintegrals. (This property is tied to the fact that integrals
(h) and (i) cannot be detected via two-particle cuts.  Integral
(d) can be detected in this way, but only by a cut that splits 
it into a product of two one-loop integrals.)
Thus, only the integrals (a), (b), (c), (e), (f) and
(g) in \fig{IntegralsThreeLoopFigure} contribute to the
divergence in odd dimensions.  
The first odd dimension in which a
divergence appears in the three-loop amplitude is $D=7$, 
because this is the dimension where the
first two-loop subdivergence appears~\cite{BDDPR}.

To evaluate the contribution of the two-loop subdivergence, in
principle we need the divergent parts of the two-loop planar and
nonplanar double-box integrals with two off-shell external legs.
However, a simple observation allows us to obtain the desired result
using only double-box integrals with all massless legs.  Indeed, at
the level of the singular terms in $D\ge7$, the difference between the
massive (off-shell)
and massless integrals are terms proportional to the square of
the massive momenta. When these additional terms are
inserted in the triangle graph shown in
\fig{AmplDivergenceFigure}, they cancel at least one of the
propagators of the triangle integral, leaving behind a
bubble integral with a massless momentum flowing through it.  
As explained above, such integrals vanish in dimensional regularization.

Using this observation, an efficient strategy is to combine the divergences of
the subintegrals into divergences of on-shell subamplitudes. This is
done simply by grouping together the remaining one-loop integrals with
the same remaining propagators, which is effectively done in
\fig{AmplDivergenceFigure}.  With this reorganization we directly
evaluate the divergence in \fig{AmplDivergenceFigure}(a), with the two
internal legs, 5 and 6, placed on shell.  All other contributions vanish.
The contact term vertex represents the divergence of the on-shell
two-loop four-point amplitude.  These divergences have been computed
in ref.~\cite{BDDPR} and are given by,
\begin{eqnarray}
{\cal M}_4^{\twoloop,\ D=7-2\e}\vert_{\rm pole} 
 &=& 
 {1\over2\e\ (4\pi)^7} {\pi\over3} (s_{12}^2+s_{13}^2+s_{14}^2) \, 
\times 
\left( {\kappa \over 2}\right)^6 
\times s_{12} s_{13} s_{14} \, M_4^{\rm tree} 
\,, \nn \\ 
{\cal M}_4^{\twoloop,\ D=9-2\e}\vert_{\rm pole} 
 &=&  {1\over4\e\ (4\pi)^9} {-13\pi\over9072} 
 (s_{12}^2+s_{13}^2+s_{14}^2)^2 \, 
\times 
\left( {\kappa \over 2}\right)^6 \times s_{12} s_{13} s_{14}\, 
           M_4^{\rm tree} \, , \nn \\ 
{\cal M}_4^{\twoloop,\ D=11-2\e}\vert_{\rm pole} 
 &=&  {1\over48\e\ (4\pi)^{11}} {\pi\over5791500} 
  \Bigl( 438 (s_{12}^6+s_{13}^6+s_{14}^6) 
              - 53 s_{12}^2 s_{13}^2 s_{14}^2 \Bigr) \, \nn \\
&& \hskip 3 cm \null
\times 
\left( {\kappa \over 2}\right)^6 \times s_{12} s_{13} s_{14} \,
           M_4^{\rm tree} \,, 
\label{TwoLoopDiv}
\end{eqnarray}
for external legs labeled $1,2,3,4$. The factor $s_{12} s_{13} s_{14}$
cancels all kinematic poles in the tree amplitudes $M_4^{\rm tree}$,
making these divergences local.  (In fact, the product 
$s_{12} s_{13} s_{14} M_4^{\rm tree}$ in the case of four gravitons
may be represented as the matrix element on four-particle states
of a particular contraction of four Riemann tensors, often denoted 
simply by ``$R^4$''.)
We have also obtained the same result~(\ref{TwoLoopDiv})
by evaluating the two-loop subdivergences integral by integral,
using the results of ref.~\cite{BDDPR}.

To evaluate the remaining one-loop integrals, we again make use of 
the observation that if a numerator factor collapses either of
the propagators in \fig{AmplDivergenceFigure}(a), labeled by 5 and 6,
then the integral vanishes because it depends only on a massless
external momentum, either $k_3^2 = 0$ or $k_4^2 = 0$, as in
\fig{AmplDivergenceFigure}(b).  Therefore
we may set $l_5^2 = 0$ and $l_6^2=0$ in the numerators of
the integrals.  We may use the on-shell two-loop
divergence~(\ref{TwoLoopDiv}) and tree amplitude 
directly in the integration, giving us
\begin{eqnarray}
{\cal M}_4^{\threeloop, D} \Bigr|_{\rm pole} &=&
\frac{1}{2} \sum_{\rm perms} \,
\sum_{{\cal N}=8\ \rm states}\int {d^D l_5 \over (2 \pi)^D}\; 
\label{OneLoopIntegral} \\
&& \hskip 2 cm \null 
\times
\Bigl[
{\cal M}_4^{\twoloop, D}(-l_5, 1, 2, l_6) \Bigr|_{\rm
pole}\Bigr] \, {i \over l_6^2} \, 
M_4^{\tree}(-l_6, 3, 4, l_5) \, {i \over l_5^2} \,, \nn
\end{eqnarray}
where the factor of $1/2$ accounts for identical particles
crossing the cut, 
and the sum over permutations runs over the six distinct labellings of
\fig{AmplDivergenceFigure}(a).  To evaluate this integral we note that
the two-loop divergences (\ref{TwoLoopDiv}) are proportional to the
tree-level four-graviton amplitude.  Thus, we may use the two-particle
cut sewing relation for tree-level amplitudes~\cite{BDDPR},
\begin{eqnarray}
&& \hskip -.4 cm 
\sum_{{\cal N} =8 \rm\ states} M_4^{\tree}(-l_5, 1, 2, l_6) \times
  M_4^{\rm tree}(-l_6, 3, 4, l_5) \label{BasicGravityCutting} \\
&&  \hskip .7 cm \null 
= i\, s_{12} s_{13} s_{14} \, M_4^{\tree}(1, 2, 3, 4) 
 \biggl[{1\over (l_5 - k_1)^2 } + {1\over (l_5 - k_2)^2} \biggr]
\biggl[{1\over (l_6 - k_3)^2 } + {1\over (l_6 - k_4)^2} \biggr]\,.
\hskip .8 cm \nn
\end{eqnarray}
In the sewing (\ref{OneLoopIntegral}), the additional factor
$s_{12}s_{13}s_{14}$ appearing in the two-loop divergences
(\ref{TwoLoopDiv}) should be relabeled as, 
\begin{eqnarray}
s_{12}s_{13}s_{14}\ \mapsto\ s_{12}(l_5-k_1)^2(l_5-k_2)^2\,.
\end{eqnarray}
With this additional factor the first term in the bracket in
\eqn{BasicGravityCutting} simplifies as,
\begin{eqnarray}
 s_{12}(l_5-k_1)^2(l_5-k_2)^2 
 \biggl[{1\over (l_5 - k_1)^2 } + {1\over (l_5 - k_2)^2} \biggr]
&=& s_{12} [(l_5-k_2)^2 + (l_5-k_1)^2] \nn\\
&=& -s_{12}^2 \,,
\end{eqnarray}
where we used the on-shell conditions on $l_5$ and $l_6$.  Because a propagator
cancels, we are left with the expected triangle integrals in
\eqn{OneLoopIntegral}.

The masslessness of the external legs leads to further
simplifications in handling tensor triangle integrals, which
arise from additional factors of $s_{13}$ and $s_{14}$
in the two-loop divergences~(\ref{TwoLoopDiv}).
If in these factors we let 
$s_{13} \mapsto (l_5-k_1)^2 = -2l_5\cdot k_1$
and $s_{14} \mapsto (l_6+k_1)^2 = 2l_6\cdot k_1$, 
then all contractions $l_i^\mu l_j^\nu \to \eta^{\mu\nu}$
vanish, because they are proportional to $k_1^2=0$.
Hence we just need the Feynman parameter polynomials
obtained by shifting the loop momentum in the usual way.
Evaluating the integrals over the Feynman parameters in
the appropriate odd integer dimension, and combining the pieces,
we obtain the following three-loop divergences:
\begin{eqnarray}
{M}_4^{(3)}\Big|_{\rm pole}^{D=7-2\epsilon}&=&
- \frac{\pi^{5/2}}{\epsilon \, (4\pi)^{21/2}}
\left( {\kappa\over 2}\right)^8
\,\left[s_{12}s_{13}s_{14} M_4^{\rm tree}\right] 
\sum_{Z_3}\;{(-s_{12})^{5/2} \over 1440}
\left(\; 65 \, s_{12}^2
- 8 \, s_{13} s_{14} \;\right)
 \,, \nn
\\
{M}_4^{(3)}\Big|_{\rm pole}^{D=9-2\epsilon}&=&
- \frac{\pi^{5/2}}{\epsilon\,(4\pi)^{27/2}}
\left({\kappa \over 2}\right)^8
\, \left[s_{12}s_{13}s_{14} M_4^{\rm tree}\right]
\sum_{Z_3}
\frac{13\;(-s_{12})^{7/2}}{11705057280}
 \nn \\
&&
\times
\left(\; 10143\, s_{12}^4
- 1296 \, s_{12}^2 s_{13} s_{14}
+ 128 \, s_{13}^2 s_{14}^2 \;\right)
\,,
\nn \\
{M}_4^{(3)}\Big|_{\rm pole}^{D=11-2\epsilon}&=&
 - \frac{\pi^{5/2}}{\epsilon\,(4\pi)^{33/2}}
\left({\kappa \over 2}\right)^8\left[s_{12}s_{13}s_{14}
M_4^{\rm tree}\right]\sum_{Z_3}
{(-s_{12})^{9/2} \over 2461954796421120000} \nn \\
&&
\times
\left(\;
   3180433113\, s_{12}^6
 - 247667992 \, s_{12}^4 s_{13} s_{14}\right.\nonumber\\[4pt]
&&~~~~~~~~\left.\null
 + 70002816 \, s_{12}^2 s_{13}^2 s_{14}^2
 - 3363840 \, s_{13}^3 s_{14}^3 \; \right)
 \,, \label{HigherDimDivergences}
\end{eqnarray}
where $Z_3$ refers to cyclic permutations of legs 2, 3 and 4.
%
These divergences are really due to
two-loop subdivergences in the bare theory.  If we renormalize the
theory at two loops in $D=7$, 9 or 11, in order to cancel the
divergences in \eqn{TwoLoopDiv}, then the corresponding three-loop
divergence in the renormalized theory will also be cancelled.

\section{Conclusions}
\label{ConclusionSection}

Maximally supersymmetric $\Neqeight$ supergravity theory is potentially
a perturbatively ultraviolet-finite point-like 
quantum field theory of gravity.  
In ref.~\cite{GravityThree} a loop-integral representation of the
three-loop four-point amplitude of $\NeqEight$ supergravity was
presented, which exhibited cancellations beyond those needed for
finiteness.  In this paper, using the method of maximal
cuts~\cite{FiveLoop}, we constructed an alternate representation of
this amplitude with all ultraviolet cancellations manifest.  By
explicitly evaluating the integrals, we demonstrated that $\NeqEight$
supergravity~\cite{CremmerJuliaScherk} diverges in $D=6$,
matching the ultraviolet behavior of $\NeqFour$ super-Yang-Mills theory.
Hence no further hidden cancellations are present at three loops.
We found that the divergence has a uniform degree of transcendentality,
and is proportional to $\zeta_3$.  We also evaluated the 
divergence of the three-loop four-graviton 
amplitude in $7$, $9$ and $11$ dimensions; the latter results may 
be of interest in studies of M theory dualities~\cite{GreenVanhove}.

While explicit calculations in $\NeqEight$ supergravity reveal
cancellations beyond those needed for finiteness, their origin remains
to be fully unraveled. For a subset of contributions, all-loop
cancellations~\cite{Finite} follow from the ``no-triangle'' property
at one-loop~\cite{OneloopMHVGravity, NoTriangle, NoTriangleSixPt,
NoTriangleKallosh, BjerrumVanhove, NoTriangleProof, AHCKGravity}.  In
ref.~\cite{NoTri} one-loop cancellations in generic theories of
gravity were linked to unexpectedly soft behavior of tree-level
gravity amplitudes under large complex shifts of their
momenta~\cite{GravityRecursion, CachazoLargez, AHK, CheungGravity,
EFKRecursion}.  This mechanism was also proposed as a source of
all-loop cancellations, which may be
sufficiently strong to render the $\NeqEight$
theory finite, when combined with supersymmetric cancellations.
This line of reasoning has been pursued further in ref.~\cite{AHCKGravity}.
Improved ultraviolet properties in $\NeqEight$ supergravity have also
been linked to M theory dualities~\cite{DualityArguments,GOS} and to
string theory non-renormalization theorems~\cite{Berkovits,GreenII}.

We also presented the fully color-dressed $\NeqFour$ super-Yang-Mills
three-loop four-point amplitude.
In our color decomposition the contact terms
are given the same color factor as the parent diagram to which they are
assigned.  The parent diagrams contain only three-vertices, each of
which carries an $f^{abc}$ color factor.  The freedom to assign
contact terms to different parent diagrams, and thereby to different color
factors, cancels in the full amplitude.  To confirm our color dressing
for the three-loop four-point amplitudes, we evaluated the cuts using
color-dressed tree amplitudes as input.  We expect that, in general,
any amplitude in any gauge theory can be color dressed by first
assigning contact terms to parent diagrams, in a way that is consistent 
with all color-ordered unitarity cuts.  Then one dresses the three-point
vertices of the parent diagrams with the appropriate $f^{abc}$
color factors.

Representations of amplitudes manifestly exhibiting all ultraviolet
cancellations, such as the one presented in this paper, should be
helpful for studying their properties and for tracking the origin of
the cancellations, at both three and higher loop orders.


\section*{Acknowledgments}
\vskip -.3 cm 

We thank David Kosower for many helpful discussions and
collaboration on this topic. We also thank Paul Howe, Harald Ita,
Renata Kallosh, Kelly Stelle and Pierre Vanhove for valuable discussions.
We thank Academic Technology Services at UCLA for
computer support.  This research was supported by the US Department of
Energy under contracts DE--FG03--91ER40662 (Z. B., J. J. M. C., H. J.), 
DE--AC02--76SF00515 (L. J. D.), DE-FG02-90ER40577 (OJI) (R. R.), 
the US National Science Foundation under grants PHY-0455649
and PHY-0608114, and the A. P. Sloan Foundation (R. R.).
J.~J.~M.~C. and H.~J. gratefully acknowledge the financial support
of Guy Weyl Physics and Astronomy Alumni Fellowships.



\begin{thebibliography}{99}

\bibitem{CremmerJuliaScherk}
E.~Cremmer, B.~Julia and J.~Scherk,
Phys.\ Lett.\ B {\bf 76}, 409 (1978);\\
%
E.~Cremmer and B.~Julia,
Phys.\ Lett.\ B {\bf 80}, 48 (1978);
  Nucl.\ Phys.\  B {\bf 159}, 141 (1979).

\bibitem{GravityThree}
Z.~Bern, J.~J.~Carrasco, L.~J.~Dixon, H.~Johansson, D.~A.~Kosower 
and R.~Roiban,
Phys.\ Rev.\ Lett.\  {\bf 98}, 161303 (2007)
[hep-th/0702112].

\bibitem{GeneralizedUnitarity}
Z.~Bern, L.~J.~Dixon and D.~A.~Kosower,
Nucl.\ Phys.\ B {\bf 513}, 3 (1998) 
[hep-ph/9708239];
%
JHEP {\bf 0408}, 012 (2004)
[hep-ph/0404293];\\
%
Z.~Bern, V.~Del Duca, L.~J.~Dixon and D.~A.~Kosower,
Phys.\ Rev.\  D {\bf 71}, 045006 (2005)
[hep-th/0410224].

\bibitem{OneloopMHVGravity}
Z.~Bern, L.~J.~Dixon, M.~Perelstein and J.~S.~Rozowsky,
Nucl.\ Phys.\ B {\bf 546}, 423 (1999)
[hep-th/9811140].

\bibitem{NoTriangle}
Z.~Bern, N.~E.~J.~Bjerrum-Bohr and D.~C.~Dunbar,
JHEP {\bf 0505}, 056 (2005)
[hep-th/0501137].

\bibitem{NoTriangleSixPt}
N.~E.~J.~Bjerrum-Bohr, D.~C.~Dunbar and H.~Ita,
Phys.\ Lett.\  B {\bf 621}, 183 (2005)
[hep-th/0503102];\\
N.~E.~J.~Bjerrum-Bohr, D.~C.~Dunbar, H.~Ita, W.~B.~Perkins and K.~Risager,
JHEP {\bf 0612}, 072 (2006)
[hep-th/0610043].

\bibitem{NoTriangleKallosh}
R.~Kallosh,
0711.2108 [hep-th].

\bibitem{BjerrumVanhove}
N.~E.~J.~Bjerrum-Bohr and P.~Vanhove,
JHEP {\bf 0804}, 065 (2008)
[0802.0868 [hep-th]].

\bibitem{NoTriangleProof}
N.~E.~J.~Bjerrum-Bohr and P.~Vanhove,
0805.3682 [hep-th];
%
0806.1726 [hep-th].

\bibitem{AHCKGravity}
N.~Arkani-Hamed, F.~Cachazo and J.~Kaplan,
0808.1446 [hep-th].

\bibitem{Finite}
Z.~Bern, L.~J.~Dixon and R.~Roiban,
Phys.\ Lett.\ B {\bf 644}, 265 (2007)
[hep-th/0611086].

\bibitem{DualityArguments}
G.~Chalmers,
hep-th/0008162;\\
M.~B.~Green, J.~G.~Russo and P.~Vanhove,
JHEP {\bf 0702}, 099 (2007)
[hep-th/0610299].

\bibitem{Berkovits}
N.~Berkovits,
Phys.\ Rev.\ Lett.\  {\bf 98}, 211601 (2007)
[hep-th/0609006].

\bibitem{GreenII}
M.~B.~Green, J.~G.~Russo and P.~Vanhove,
Phys.\ Rev.\ Lett.\  {\bf 98}, 131602 (2007)
[hep-th/0611273].

\bibitem{GOS}
M.~B. Green, H. Ooguri and J.~H. Schwarz,
Phys.\ Rev.\ Lett.\  {\bf 99}, 041601 (2007)
[0704.0777 [hep-th]].

\bibitem{Supergravity}
M.~T.~Grisaru,
Phys.\ Lett.\  B {\bf 66}, 75 (1977);\\
%
E.~Tomboulis,
Phys.\ Lett.\  B {\bf 67}, 417 (1977);\\
%
S.~Deser, J.~H.~Kay and K.~S.~Stelle,
Phys.\ Rev.\ Lett.\  {\bf 38}, 527 (1977);\\
%
P.~S.~Howe and K.~S.~Stelle,
Int.\ J.\ Mod.\ Phys.\ A {\bf 4}, 1871 (1989);\\
%
 N.~Marcus and A.~Sagnotti,
  Nucl.\ Phys.\  B {\bf 256}, 77 (1985).

\bibitem{HoweStelleNew}
P.~S.~Howe and K.~S.~Stelle,
Phys.\ Lett.\ B {\bf 554}, 190 (2003)
[hep-th/0211279].

\bibitem{GrisaruSiegel}
M.~T.~Grisaru and W.~Siegel,
Nucl.\ Phys.\  B {\bf 201}, 292 (1982)
[Erratum-ibid.\  B {\bf 206}, 496 (1982)].

\bibitem{Kallosh}
P.~S.~Howe and U.~Lindstrom,
Nucl.\ Phys.\  B {\bf 181}, 487 (1981);\\
%
R.~E.~Kallosh,
Phys.\ Lett.\  B {\bf 99}, 122 (1981).

\bibitem{KellyPrivate}
K.~S.~Stelle, talk presented at the 
{\it UCLA workshop: Is $\NeqEight$ Supergravity Finite?},
http://www.physics.ucla.edu/tep/\~{}workshops/supergravity/agenda/Talks/talk2.pdf

\bibitem{NoTri}
Z.~Bern, J.~J.~Carrasco, D.~Forde, H.~Ita and H.~Johansson,
Phys.\ Rev.\  D {\bf 77}, 025010 (2008)
[0707.1035 [hep-th]].

\bibitem{BCFRecursion}
R.~Britto, F.~Cachazo and B.~Feng,
Nucl.\ Phys.\  B {\bf 715}, 499 (2005)
[hep-th/0412308].

\bibitem{BCFW}
R.~Britto, F.~Cachazo, B.~Feng and E.~Witten,
Phys.\ Rev.\ Lett.\  {\bf 94}, 181602 (2005)
[hep-th/0501052].

\bibitem{GravityRecursion}
J.~Bedford, A.~Brandhuber, B.~J.~Spence and G.~Travaglini,
Nucl.\ Phys.\  B {\bf 721}, 98 (2005)
[hep-th/0502146];\\
%
F.~Cachazo and P.~Svr\v{c}ek,
hep-th/0502160;\\
%
N.~E.~J.~Bjerrum-Bohr, D.~C.~Dunbar, H.~Ita, W.~B.~Perkins and K.~Risager,
JHEP {\bf 0601}, 009 (2006)
[hep-th/0509016];\\
%
A.~Brandhuber, S.~McNamara, B.~Spence and G.~Travaglini,
JHEP {\bf 0703}, 029 (2007)
[hep-th/0701187].

\bibitem{CachazoLargez}
P.~Benincasa, C.~Boucher-Veronneau and F.~Cachazo,
JHEP {\bf 0711}, 057 (2007)
[hep-th/0702032];\\
%
P.~Benincasa and F.~Cachazo,
0705.4305 [hep-th];\\
%
A.~Hall,
Phys.\ Rev.\  D {\bf 77}, 124004 (2008)
[0803.0215 [hep-th]].

\bibitem{EFKRecursion}
H.~Elvang, D.~Z.~Freedman and M.~Kiermaier,
0808.1720 [hep-th].

\bibitem{AHK}
N.~Arkani-Hamed and J.~Kaplan,
JHEP {\bf 0804}, 076 (2008)
[0801.2385 [hep-th]].

\bibitem{CheungGravity}
C.~Cheung,
0808.0504 [hep-th].

\bibitem{SpaceCone}
G.~Chalmers and W.~Siegel,
Phys.\ Rev.\  D {\bf 59}, 045013 (1999)
[hep-ph/9801220];\\
%
D.~Vaman and Y.~P.~Yao,
JHEP {\bf 0604}, 030 (2006)
[hep-th/0512031];
%
0805.2645 [hep-th].

\bibitem{tHooftVeltmanGravity}
G.~'t Hooft and M.~J.~G.~Veltman,
Annales Poincare Phys.\ Theor.\ A {\bf 20}, 69 (1974).

\bibitem{DeserMatter}
S.~Deser and P.~van Nieuwenhuizen,
Phys.\ Rev.\  D {\bf 10}, 401 (1974);\\
%
S.~Deser, H.~S.~Tsao and P.~van Nieuwenhuizen,
Phys.\ Rev.\  D {\bf 10}, 3337 (1974).

\bibitem{DunbarNorridge}
D.~C.~Dunbar and P.~S.~Norridge,
Class.\ Quant.\ Grav.\  {\bf 14}, 351 (1997)
[hep-th/9512084];\\
%
D.~C.~Dunbar and P.~S.~Norridge,
Nucl.\ Phys.\  B {\bf 433}, 181 (1995)
[hep-th/9408014].

\bibitem{Kallosh74}
R.~E.~Kallosh,
Nucl.\ Phys.\  B {\bf 78}, 293 (1974).

\bibitem{vanNWu}
P.~van Nieuwenhuizen and C.~C.~Wu,
J.\ Math.\ Phys.\  {\bf 18}, 182 (1977).

\bibitem{GoroffSagnotti}
M.~H.~Goroff and A.~Sagnotti,
Phys.\ Lett.\  B {\bf 160}, 81 (1985);
Nucl.\ Phys.\  B {\bf 266}, 709 (1986).
%

\bibitem{vandeVen}
A.~E.~M.~van de Ven,
Nucl.\ Phys.\ B {\bf 378}, 309 (1992).

\bibitem{UnitarityMethod}
Z.~Bern, L.~J.~Dixon, D.~C.~Dunbar and D.~A.~Kosower,
Nucl.\ Phys.\ B {\bf 425}, 217 (1994)
[hep-ph/9403226];\\
%
Z.~Bern, L.~J.~Dixon, D.~C.~Dunbar and D.~A.~Kosower,
Nucl.\ Phys.\ B {\bf 435}, 59 (1995)
[hep-ph/9409265].

\bibitem{BDDPR}
Z.~Bern, L.~J.~Dixon, D.~C.~Dunbar, M.~Perelstein and J.~S.~Rozowsky,
Nucl.\ Phys.\ B {\bf 530}, 401 (1998)
[hep-th/9802162].

\bibitem{BRY}
Z.~Bern, J.~S.~Rozowsky and B.~Yan,
Phys.\ Lett.\  B {\bf 401}, 273 (1997)
[hep-ph/9702424].

\bibitem{KLT}
H.~Kawai, D.~C.~Lewellen and S.~H.~H.~Tye,
Nucl.\ Phys.\ B {\bf 269}, 1 (1986).

\bibitem{KLT2}
Z.~Bern and A.~K.~Grant,
Phys.\ Lett.\  B {\bf 457}, 23 (1999)
[hep-th/9904026];\\
%
Z.~Bern, A.~De Freitas and H.~L.~Wong,
Phys.\ Rev.\ Lett.\  {\bf 84}, 3531 (2000)
[hep-th/9912033];\\
%
N.~E.~J.~Bjerrum-Bohr,
Phys.\ Lett.\  B {\bf 560}, 98 (2003)
[hep-th/0302131];
%
Nucl.\ Phys.\  B {\bf 673}, 41 (2003)
[hep-th/0305062];\\
%
N.~E.~J.~Bjerrum-Bohr and K.~Risager,
Phys.\ Rev.\  D {\bf 70}, 086011 (2004)
[hep-th/0407085];\\
%
S. Ananth and S. Theisen,
Phys.\ Lett.\  B {\bf 652}, 128 (2007)
[0706.1778 [hep-th]];\\
%
H.~Elvang and D.~Z.~Freedman,
JHEP {\bf 0805}, 096 (2008)
[0710.1270 [hep-th]].

\bibitem{GravityReview}
Z.~Bern,
Living Rev.\ Rel.\  {\bf 5}, 5 (2002)
[gr-qc/0206071].

\bibitem{BEZ}
M.~Bianchi, H.~Elvang and D.~Z.~Freedman,
0805.0757 [hep-th].

\bibitem{TreeJacobi}
Z.~Bern, J.~J.~M.~Carrasco and H.~Johansson,
0805.3993 [hep-ph].

\bibitem{FiveLoop}
Z.~Bern, J.~J.~M.~Carrasco, H.~Johansson and D.~A.~Kosower,
Phys.\ Rev.\  D {\bf 76}, 125020 (2007)
[0705.1864 [hep-th]].

\bibitem{BCFGeneralized}
R.~Britto, F.~Cachazo and B.~Feng,
Nucl.\ Phys.\  B {\bf 725}, 275 (2005)
[hep-th/0412103];\\
%
E.~I.~Buchbinder and F.~Cachazo,
JHEP {\bf 0511}, 036 (2005)
[hep-th/0506126].

\bibitem{WittenTopologicalString}
E.~Witten,
Commun.\ Math.\ Phys.\  {\bf 252}, 189 (2004)
[hep-th/0312171].

\bibitem{FreddyMaximal}
F.~Cachazo and D.~Skinner,
0801.4574 [hep-th];\\
%
F.~Cachazo,
0803.1988 [hep-th].

\bibitem{LeadingSingularityCalcs}
F.~Cachazo, M.~Spradlin and A.~Volovich,
0805.4832 [hep-th];\\
%
M.~Spradlin, A.~Volovich and C.~Wen,
0808.1054 [hep-th].

\bibitem{FDH}
Z.~Bern and D.~A.~Kosower,
Nucl.\ Phys.\  B {\bf 379}, 451 (1992);\\
%
Z.~Bern, A.~De Freitas, L.~J.~Dixon and H.~L.~Wong,
Phys.\ Rev.\  D {\bf 66}, 085002 (2002)
[hep-ph/0202271].

\bibitem{DimRed}
W.~Siegel,
Phys.\ Lett.\  B {\bf 84}, 193 (1979).

\bibitem{Nair}
V.~P.~Nair,
Phys.\ Lett.\  B {\bf 214}, 215 (1988).

\bibitem{Witten_twistor}
E.~Witten,
Commun.\ Math.\ Phys.\  {\bf 252}, 189 (2004)
[hep-th/0312171].

\bibitem{GGK}
G.~Georgiou, E.~W.~N.~Glover and V.~V.~Khoze,
JHEP {\bf 0407}, 048 (2004)
[hep-th/0407027].

\bibitem{RecentOnShellSuperSpace}
J.~M.~Drummond, J.~Henn, G.~P.~Korchemsky and E.~Sokatchev,
0807.1095 [hep-th];\\
%
A.~Brandhuber, P.~Heslop and G.~Travaglini,
0807.4097 [hep-th];\\
%
J.~M.~Drummond, J.~Henn, G.~P.~Korchemsky and E.~Sokatchev,
0808.0491 [hep-th];\\
%
J.~M.~Drummond and J.~M.~Henn,
0808.2475 [hep-th].

\bibitem{GreenVanhove}
M.~B.~Green, M.~Gutperle and P.~Vanhove,
Phys.\ Lett.\  B {\bf 409}, 177 (1997)
[hep-th/9706175];\\
%
M.~B.~Green, H.~h.~Kwon and P.~Vanhove,
Phys.\ Rev.\  D {\bf 61}, 104010 (2000)
[hep-th/9910055];\\
%
M.~B.~Green and P.~Vanhove,
JHEP {\bf 0601}, 093 (2006)
[hep-th/0510027];\\
%
M.~B.~Green, J.~G.~Russo and P.~Vanhove,
JHEP {\bf 0802}, 020 (2008)
[0801.0322 [hep-th]];
%
JHEP {\bf 0807}, 126 (2008)
[0807.0389 [hep-th]].

\bibitem{IRBehavior}
R.~Akhoury,
Phys.\ Rev.\  D {\bf 19}, 1250 (1979);\\
%
A.~H.~Mueller,
Phys.\ Rev.\  D {\bf 20}, 2037 (1979);\\
%
J.~C.~Collins,
Phys.\ Rev.\  D {\bf 22}, 1478 (1980);\\
%
A.~Sen,
Phys.\ Rev.\  D {\bf 24}, 3281 (1981);\\
%
G.~Sterman,
Nucl.\ Phys.\  B {\bf 281}, 310 (1987);\\
%
J.~Botts and G.~Sterman,
Phys.\ Lett.\  B {\bf 224}, 201 (1989)
[Erratum-ibid.\  B {\bf 227}, 501 (1989)];\\
%
S.~Catani and L.~Trentadue,
Nucl.\ Phys.\  B {\bf 327}, 323 (1989);\\
%
G.~P.~Korchemsky,
Phys.\ Lett.\  B {\bf 220}, 629 (1989);\\
%
L.~Magnea and G.~Sterman,
Phys.\ Rev.\  D {\bf 42}, 4222 (1990);\\
%
G.~P.~Korchemsky and G.~Marchesini,
Phys.\ Lett.\  B {\bf 313} (1993) 433;\\
%
S.~Catani,
Phys.\ Lett.\  B {\bf 427}, 161 (1998)
[hep-ph/9802439];\\
%
G.~Sterman and M.~E.~Tejeda-Yeomans,
Phys.\ Lett.\  B {\bf 552}, 48 (2003)
[hep-ph/0210130];\\
%
L.~J.~Dixon, L.~Magnea and G.~Sterman,
JHEP {\bf 0808}, 022 (2008)
[0805.3515 [hep-ph]].

\bibitem{SoftMatrix}
A.~Sen,
Phys.\ Rev.\  D {\bf 28}, 860 (1983);\\
%
J.~Botts and G.~Sterman,
Nucl.\ Phys.\  B {\bf 325}, 62 (1989);\\
%
N.~Kidonakis, G.~Oderda and G.~Sterman,
Nucl.\ Phys.\  B {\bf 525}, 299 (1998)
[hep-ph/9801268];
%
Nucl.\ Phys.\  B {\bf 531}, 365 (1998)
[hep-ph/9803241];\\
%
A.~Banfi, G.~P.~Salam and G.~Zanderighi,
JHEP {\bf 0408}, 062 (2004)
[hep-ph/0407287].

\bibitem{AybatDixonSterman}
S.~Mert Aybat, L.~J.~Dixon and G.~Sterman,
Phys.\ Rev.\ Lett.\  {\bf 97}, 072001 (2006)
[hep-ph/0606254];
%
Phys.\ Rev.\  D {\bf 74}, 074004 (2006)
[hep-ph/0607309].

\bibitem{KorchemskyRadyushkin}
G.~P.~Korchemsky and A.~V.~Radyushkin,
Phys.\ Lett.\ B {\bf 171} (1986) 459;\\
%
S.~V.~Ivanov, G.~P.~Korchemsky and A.~V.~Radyushkin,
Yad.\ Fiz.\ {\bf 44} (1986) 230
[Sov.\ J.\ Nucl.\ Phys.\ {\bf 44} (1986) 145].

\bibitem{BDS}
Z.~Bern, L.~J.~Dixon and V.~A.~Smirnov,
Phys.\ Rev.\ D {\bf 72}, 085001 (2005)
[hep-th/0505205].

\bibitem{SmirnovTripleBox}
V.~A.~Smirnov,
Phys.\ Lett.\  B {\bf 567}, 193 (2003)
[hep-ph/0305142].

\bibitem{ABDK}
C.~Anastasiou, Z.~Bern, L.~J.~Dixon and D.~A.~Kosower,
Phys.\ Rev.\ Lett.\  {\bf 91}, 251602 (2003)
[hep-th/0309040].

\bibitem{Iterate}
F.~Cachazo, M.~Spradlin and A.~Volovich,
Phys.\ Rev.\  D {\bf 74}, 045020 (2006)
[hep-th/0602228];\\
%
Z.~Bern, M.~Czakon, D.~A.~Kosower, R.~Roiban and V.~A.~Smirnov,
Phys.\ Rev.\ Lett.\  {\bf 97}, 181601 (2006)
[hep-th/0604074];\\
%
L.~F.~Alday and J.~Maldacena,
JHEP {\bf 0706}, 064 (2007)
[0705.0303 [hep-th]];
%
JHEP {\bf 0711}, 068 (2007)
[0710.1060 [hep-th]];\\
%
J.~M.~Drummond, J.~Henn, G.~P.~Korchemsky and E.~Sokatchev,
0712.1223 [hep-th];\\
%
J.~Bartels, L.~N.~Lipatov and A.~S.~Vera,
0802.2065 [hep-th];
%
0807.0894 [hep-th];\\
%
Z.~Bern, L.~J.~Dixon, D.~A.~Kosower, R.~Roiban, M.~Spradlin, C.~Vergu
and A.~Volovich,
Phys.\ Rev.\  D {\bf 78}, 045007 (2008)
[0803.1465 [hep-th]].

\bibitem{Weinberg}
S.~Weinberg,
Phys.\ Rev.\  {\bf 140}, B516 (1965).

\bibitem{GravityIR}
S.~G.~Naculich, H.~Nastase and H.~J.~Schnitzer,
0805.2347 [hep-th];\\
%
A.~Brandhuber, P.~Heslop, A.~Nasti, B.~Spence and G.~Travaglini,
0805.2763 [hep-th].

\bibitem{E7Original}
M.~K.~Gaillard and B.~Zumino,
Nucl.\ Phys.\  B {\bf 193}, 221 (1981);\\
%
B.~de Wit and H.~Nicolai,
Nucl.\ Phys.\  B {\bf 208}, 323 (1982).

\bibitem{E7Recent}
L.~Brink, S.~S.~Kim and P.~Ramond,
JHEP {\bf 0806}, 034 (2008)
[0801.2993 [hep-th]];\\
R.~Kallosh and M.~Soroush,
Nucl.\ Phys.\  B {\bf 801}, 25 (2008)
[0802.4106 [hep-th]].

\bibitem{HoweStelleYangMills}
P.~S.~Howe and K.~S.~Stelle,
Phys.\ Lett.\ B {\bf 137}, 175 (1984).

\bibitem{HarmonicSuperspace}
A.~Galperin, E.~Ivanov, S.~Kalitsyn, V.~Ogievetsky and E.~Sokatchev,
Class.\ Quant.\ Grav.\  {\bf 1}, 469 (1984);
%
Class.\ Quant.\ Grav.\  {\bf 2}, 155 (1985).

\bibitem{DDimUnitarity}
Z.~Bern and A.~G.~Morgan,
Nucl.\ Phys.\ B {\bf 467}, 479 (1996)
[hep-ph/9511336];
%
Z.~Bern, L.~J.~Dixon and D.~A.~Kosower,
Ann.\ Rev.\ Nucl.\ Part.\ Sci.\  {\bf 46}, 109 (1996)
[hep-ph/9602280];\\
%
Z.~Bern, L.~J.~Dixon, D.~C.~Dunbar and D.~A.~Kosower,
Phys.\ Lett.\  B {\bf 394}, 105 (1997)
[hep-th/9611127];\\
%
Z.~Bern, L.~J.~Dixon and D.~A.~Kosower,
JHEP {\bf 0001}, 027 (2000)
[hep-ph/0001001].

\bibitem{SpinorHelicity}
F.~A.~Berends, R.~Kleiss, P.~De Causmaecker, R.~Gastmans and T.~T.~Wu,
Phys.\ Lett.\ B {\bf 103}, 124 (1981);\\
%
P.~De Causmaecker, R.~Gastmans, W.~Troost and T.~T.~Wu,
Nucl.\ Phys.\ B {\bf 206}, 53 (1982);\\
%
Z.~Xu, D.~H.~Zhang and L.~Chang,
TUTP-84/3-TSINGHUA;\\
%
R.~Kleiss and W.~J.~Stirling,
Nucl.\ Phys.\ B {\bf 262}, 235 (1985);\\
%
J.~F.~Gunion and Z.~Kunszt,
Phys.\ Lett.\ B {\bf 161}, 333 (1985);\\
%
Z.~Xu, D.~H.~Zhang and L.~Chang,
Nucl.\ Phys.\ B {\bf 291}, 392 (1987).

\bibitem{LanceColor}
V.~Del Duca, A.~Frizzo and F.~Maltoni,
Nucl.\ Phys.\ B{\bf 568},  211 (2000)
[hep-ph/9909464];\\
%
V.~Del Duca, L.~J.~Dixon and F.~Maltoni,
Nucl.\ Phys.\  B {\bf 571}, 51 (2000)
[hep-ph/9910563].

\bibitem{KleissKuijf}
R.~Kleiss and H.~Kuijf,
Nucl.\ Phys.\  B {\bf 312}, 616 (1989).

\bibitem{GrozinLectures}
A.~G.~Grozin,
Int.\ J.\ Mod.\ Phys.\  A {\bf 19}, 473 (2004)
[hep-ph/0307297].

\bibitem{IBP}
F.V.~Tkachov,
Phys.\ Lett.\ B {\bf 100}, 65 (1981); \\
%
K.G.~Chetyrkin and F.V.~Tkachov,
Nucl.\ Phys.\ B {\bf 192}, 159 (1981).


\end{thebibliography}
\end{document}